\documentclass[journal]{IEEEtran}

\usepackage{graphicx}
\usepackage{amssymb}
\usepackage{amsfonts}
\usepackage{amsmath}
\usepackage{epsfig}
\usepackage{color}
\usepackage{fancybox}
\usepackage{textcomp}
\usepackage{multirow}
\usepackage{setspa ce}
\usepackage{psfrag}
\usepackage[ruled,vlined,linesnumbered]{algorithm2e}

\usepackage{mathrsfs}

\usepackage{booktabs}

\usepackage{cuted}

\usepackage{hyperref}

    \def\Complex{{\rm\rule[.23ex]{.03em}{1.1ex}\kern-.3em{C}}}

    \newcommand{\be}{\begin{equation}} \newcommand{\ee}{\end{equation}}
    \newcommand{\bea}{\begin{eqnarray}} \newcommand{\eea}{\end{eqnarray}}
    \newcommand{\benum}{\begin{enumerate}} \newcommand{\eenum}{\end{enumerate}}

    \newcommand{\qb}{{\bf b}}

    \newcommand{\qs}{{\bf s}}

    \newcommand{\qB}{{\bf B}}

    \newcommand{\qE}{{\bf E}}

    \newcommand{\qM}{{\bf M}}

    \newcommand{\qR}{{\bf R}}
    
    \newcommand{\qT}{{\bf T}}

    \newcommand{\bbC}{{\mathbb C}}

    \newcommand{\sfE}{{\sf E}}

    \newcommand{\sfqE}{\boldsymbol{\sf E}}

\begin{document}

\title{A Novel Channel Model for Reconfigurable Intelligent Surfaces with Consideration of Polarization and Switch Impairments}


\author{De-Ming Chian, \IEEEmembership{Graduate Student Member, IEEE},~Chao-Kai~Wen, \IEEEmembership{Senior Member, IEEE}, \\
~Chi-Hung~Wu,~Fu-Kang~Wang, \IEEEmembership{Member, IEEE}, and~Kai-Kit~Wong, \IEEEmembership{Fellow, IEEE}
\thanks{D.-M.~Chian and C.-K.~Wen are with the Institute of Communications Engineering, National Sun Yat-sen University, Kaohsiung 804, Taiwan (e-mail: $\rm icefreeman123@gmail.com$, $\rm chaokai.wen@mail.nsysu.edu.tw$).}
\thanks{C.-H.~Wu and F.-K.~Wang are with the Department of Electrical Engineering, National Sun Yat-sen University, Kaohsiung 804, Taiwan (e-mail: $\rm yes121335@gmail.com$, $\rm fkw@mail.ee.nsysu.edu.tw$).}
\thanks{{K.-K.~Wong} is with Department of Electronic and Electrical Engineering, University College London, UK, Email: {\rm kai-kit.wong@ucl.ac.uk}.}
}


\maketitle

\begin{abstract}
Future wireless networks require the ability to actively adjust the wireless environment to meet strict performance indicators. Reconfigurable Intelligent Surface (RIS) technology is gaining attention for its advantages of low power consumption, cost-effectiveness, and ease of deployment. However, existing channel models for RIS often ignore important properties, such as the impairment in the RIS's switch component and the polarization efficiency among antennas, limiting their practical use. In this paper, we propose a new channel model for RIS that considers these ignored properties, including the reflected field, scattered field, and antenna resonant mode. We verify the proposed model through practical implementation of a 4 $\times$ 4 RIS array with patch antennas in the 3.5 GHz band, using a phase shifter as the switch component of a RIS element. The equivalent model of the phase shifter is also formulated and incorporated into the channel model. We propose a blind controlling algorithm to discuss the properties of our channel model and emphasize the importance of considering polarization and tracking mechanisms for the controlling algorithm. Our channel model is an improvement over existing models and can be used in the practical design of RIS technology. The proposed algorithm provides a practical approach to controlling the wireless environment, suitable for various wireless applications.
\end{abstract}

\begin{IEEEkeywords}
Reconfigurable intelligent surface, intelligent reflecting surface, channel model, antenna polarization, digital phase shifter model, blind controlling algorithm.
\end{IEEEkeywords}

\section*{I. Introduction}

Optimizing the user experience in all scenarios is a main goal for 6G networks, which requires significant improvements in key performance indicators such as spectrum efficiency, capacity, and reliability. Achieving these objectives necessitates advanced transmitter and receiver designs, and the ability to actively adjust the wireless environment. The latest version of the 3rd Generation Partnership Project (3GPP), Release 18, is considering technologies such as Integrated Access and Backhaul, as well as Network Controlled Repeaters, to achieve these goals. Reconfigurable Intelligent Surface (RIS) technology, due to its potential benefits, such as low power consumption, cost-effectiveness, and easy deployment, is expected to be included in the next release version \cite{Yuan-2022-COMMag}.

Despite the potential benefits of RIS technology, understanding the channel model for RIS remains a persistent challenge, and limited experimentation and measurement in this area have been reported in the current literature \cite{Huang-2022}.
The channel models for RIS generally have two categories: statistical models and ray-tracing based models \cite{Yuan-2022}.  Statistical models, such as those found in \cite{Basar-2021,Dang-2021,Sun-2021}, are based on the 3GPP cluster channel model and describe the distribution of simulated channels for RIS as a controllable scattering cluster. Ray-tracing based models, such as those described in \cite{Liu-2022,TechRxiv-2022}, use the complex radar cross section of RIS's scattered field and import RIS into a ray-tracer as a secondary transmitter.

In the existing literature, RIS has mainly been regarded as a scatterer in a wireless environment. Theoretical scattered field channel models for RIS have been formulated using Green functions \cite{Pei-2021} or vector potentials \cite{AEE-BOOK,Physics-2020}. However, these models do not consider practical design approaches for RIS, leading to oversimplified channel models and naive formulating methods. As pointed out by \cite{Huang-2022}, these models have limited applications in practical measurements.

When electromagnetic (EM) waves interact with RIS, they interact with two types of material interfaces: metal and substrate, which can be designed as the two types of EM controlling methods for RIS. The first EM controlling method for RIS utilizes the antenna resonant mode corresponding to the resonant length on the metal plate. This resonant phenomenon is based on the surface current distribution \cite{ANT-BOOK,Wong-2017}. Green functions \cite{Pei-2021} and vector potentials \cite{AEE-BOOK,Physics-2020} are also based on surface currents, but they do not account for other EM phenomena. Specifically, when an incident wave impinges on the metal plate, it generates not only reflected and scattered fields but also \emph{antenna resonant modes}. For example, passive and active RIS elements \cite{Ren-2022,Pei-2021,Dunna-2020,Rao-2022,Kishor-2012} are often made of metal plates with special shapes connected by switch components, which change the resonant length and generate different antenna resonant modes. The metasurface approaches \cite{Darvazehban-2019,Darvazehban-2020,Zhang-2022-TAP} use the small resonating structures with the small-size antenna technologies \cite{Lu-2015,Lu-2016}, and are controlled by switch components based on the metal plate's antenna resonant mode.

The second EM controlling method for RIS uses the controllable properties of substrate material, and the current technology for this method is based on liquid crystals (LCs) \cite{Perez-2013,Li-2021,Guirado-2022,Kim-2022}. LCs respond to an applied electric field by changing their material permittivity \cite{Perez-2013}, and can be used to directly control the incident, reflected, and scattered fields by modifying the electric field of the incident wave passing through them. Additionally, LCs can control the coupling effect between metal plates primarily attributed to the electric fields along the air-dielectric interface by altering the surface current and resonant length of the antenna resonant mode \cite{ANT-BOOK}. However, designing and controlling LCs is challenging, and thus this technology is not yet widely used in RIS.

Besides the issue of the lack of consideration of practical design approaches, two main issues are not fully addressed in the current literature. Firstly, the phase-dependent attenuation caused by the switch components of RIS is not considered. Most current channel models for RIS assume that attenuations in different phases are equal. However, the controlled switch components of RIS, such as PIN diodes \cite{Ren-2022,Rao-2022}, varactor diodes \cite{Pei-2021}, or RF switches \cite{Dunna-2020}, switch the RIS to different phases, resulting in different attenuations. While some current works \cite{Abe-2020,Zhao-2022,Camana-2022,Zhu-2013} are gradually addressing this issue, physical phenomena of the switch components and the causes of phase-dependent attenuation are not clearly explained, and \cite{Zhu-2013} indicates that simulations based on their model still differ from measurements.

Secondly, the polarization efficiency is not given sufficient emphasis. The level of matched polarization among antennas of the transmitter, receiver, and RIS significantly affects the received power. However, current channel models for RIS either ignore polarization or only consider individual antenna gain.

It is evident from the above discussions that a new channel model is necessary to account for the practical design aspects of RIS. Furthermore, the controlling algorithm for RIS must be aligned with its corresponding channel model to ensure optimal performance, otherwise, it may not work effectively. Current algorithms, such as those in \cite{Abe-2020,Dunna-2020,Huang-2021,Zhang-2021}, rely on explicit channel state information (CSI), but acquiring this information for RIS poses significant challenges in engineering practice. These challenges include the easily overwhelmed channel of RIS by background noise, the need to modify current networking protocols, and the additional pilot overheads for channel estimation. Therefore, a controlling algorithm with implicit CSI is more suitable for practical use. To address these issues, we propose a model that considers practical RIS design and develop a blind controlling algorithm. We thoroughly evaluate the proposed model through experiments and simulations.

The main contributions of this work are as follows:
\begin{itemize}
\item A \textit{digital phase shifter (DPS) model for RIS} is proposed, based on the transmission-line model, which illustrates the cause of phase-dependent attenuation due to constructive and destructive interferences in the transmission line.

\item A \textit{practical design-aware channel model for RIS} is proposed, which considers both the level of matched polarization among antennas and the orientation of the antenna. This model provides a unique perspective on RIS by considering the reflected field, scattered field, and antenna resonant mode simultaneously, and illustrates how these three fields can be integrated into the channel model with the DPS model.

\item A \textit{blind controlling algorithm for RIS with implicit CSI} is proposed, which is suitable for practical use and demonstrates significant performance compared to other advanced approaches that do not consider the issues of phase-dependent attenuation and polarization efficiency. The algorithm's tracking capability is also evaluated when the receiver is moved.

\end{itemize}

The remainder of this article is structured as follows: In Section II, we present the DPS model and extend it to the radar cross section of RIS. In Section III, we present the practical channel model for RIS and validate it. In Section IV, we describe the blind controlling algorithm for RIS. Section V evaluates the algorithm and examines the properties of the proposed channel model. Finally, in Section VI, we conclude the paper.

\section*{II. RIS Modeling}

\begin{figure}
    \centering
    \resizebox{3.6in}{!}{%
    \includegraphics*{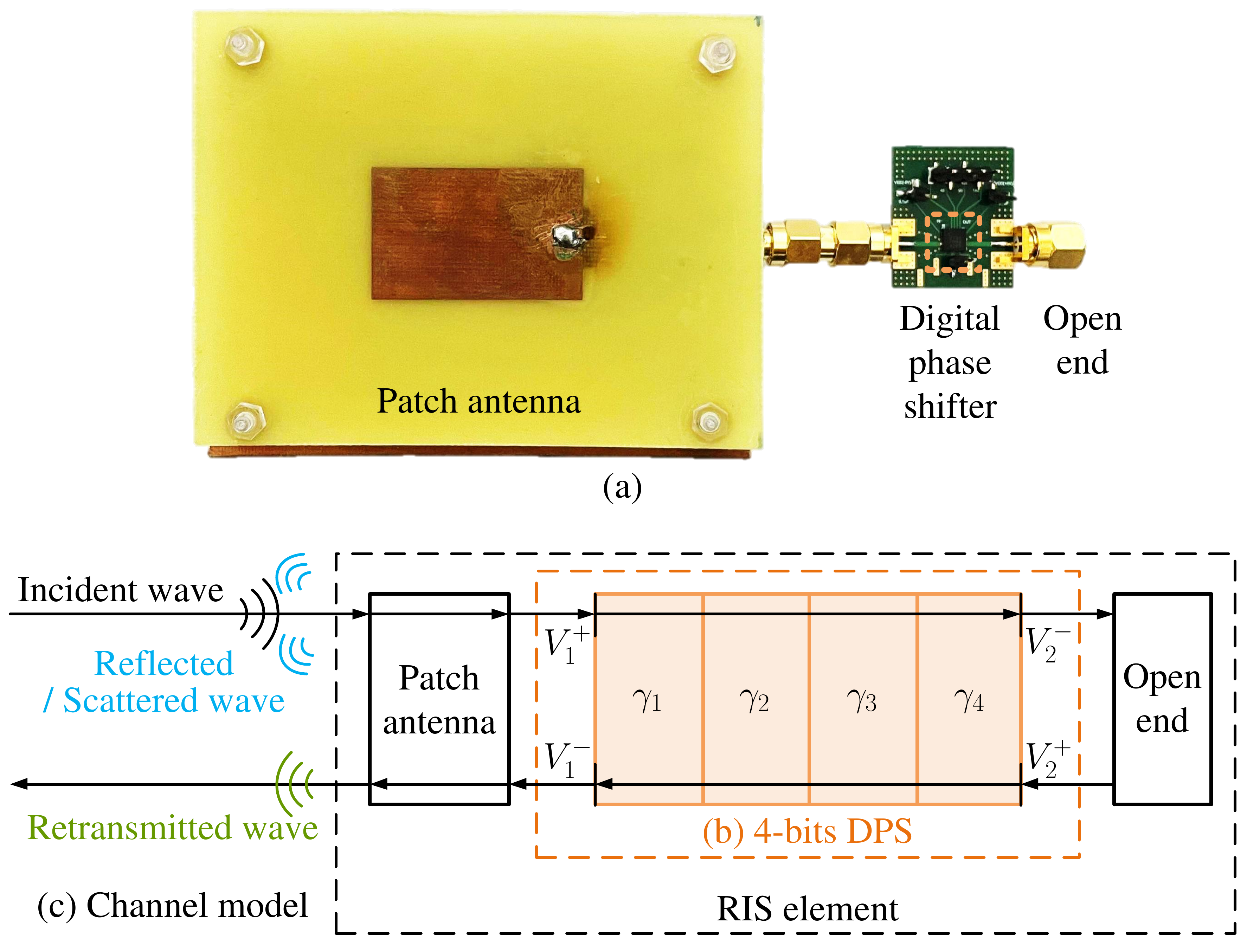} }%
    \caption{(a) RIS element, (b) transmission line model for a $4$-bits DPS in a RIS element, and (c) channel model for a RIS element.
    \label{fig:RISelement}}
\end{figure}

The structure of the RIS element is depicted in Fig. \ref{fig:RISelement}(a), it is composed of a patch antenna, a digital phase shifter (DPS), and an open end. To account for practical design considerations, we describe the design of the RIS element and its corresponding hardware model in Section II.A. Additionally, in Section II.B, we clarify the distinctions between the RIS as a scatterer and as a radiator, as many previous works (e.g., \cite{Wang-2021}) have not clearly addressed this aspect.

\subsection*{A. Digital Phase Shifter Model}

We use MAPS-010144 produced by M/A-COM Technology Solutions Inc., which is a 4-bit DPS providing a phase shift from $0^{\circ}$ to $360^{\circ}$ in $22.5^{\circ}$ steps. Compared to other switch components \cite{Ren-2022,Rao-2022,Pei-2021,Dunna-2020}, DPS has the ability to easily achieve many stable states. Thus, we can easily analyze the different states of RIS with DPS in detail. We refer to the functional schematic of MAPS-010144 and the equivalent circuit model of a multi-bit phase shifter using multiple cascaded $1$-bit phase shifters in \cite{Lee-2015}.

Assuming that the internal DPS has perfect impedance matching, the reflection coefficient of the internal DPS can be ignored. As illustrated in Fig. \ref{fig:RISelement}(b), we use the transmission-line model with S parameters \cite{RF-BOOK} to model the DPS in an RIS element as follows:
\begin{equation} \label{eq: Sparameters}
    \left[    \begin{array}{c}
    V_{1}^{-} \\
    V_{2}^{-}
    \end{array}    \right]
     =
    \left[    \begin{array}{cc}
    S_{11}^{\rm dps} & S_{12}^{\rm dps} \\
    S_{21}^{\rm dps} & S_{22}^{\rm dps}
    \end{array}     \right]
    \left[    \begin{array}{c}
    V_{1}^{+} \\
    V_{2}^{+}
    \end{array}    \right],
\end{equation}
where superscripts $^{+}$ and $^{-}$ of $V$ represent the signal going towards and travelling away from the port, respectively, and subscripts $_{1}$ and $_{2}$ of $V$ represent the signal at the two ports.
Given that the reflection coefficient of the open end is ${\Gamma_{\rm end} = V_{2}^{+}/V_{2}^{-}}$,
the reflection coefficient of the DPS with an open end (DPS-O) can be obtained by solving \eqref{eq: Sparameters}:
\begin{equation} \label{eq: DPSmodel_Spar}
    \Gamma = \frac{V_{1}^{-}}{V_{1}^{+}}
    =
    S_{11}^{\rm dps} + \frac{S_{12}^{\rm dps} \Gamma_{\rm end} S_{21}^{\rm dps}}{1 - \Gamma_{\rm end} S_{22}^{\rm dps}}.
\end{equation}

\begin{figure}
    \centering
    \resizebox{3.6in}{!}{%
    \includegraphics*{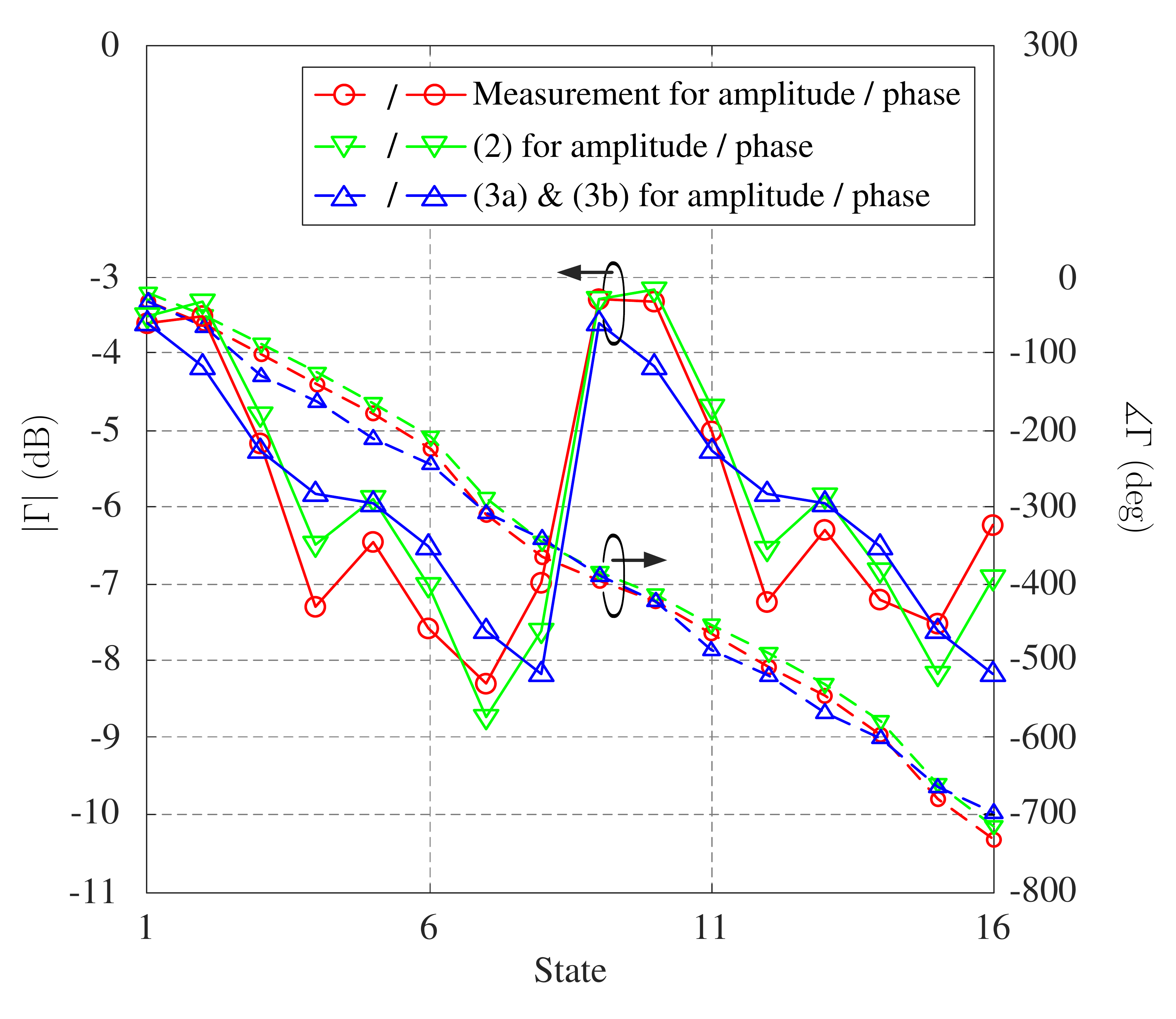} }%
    \caption{Reflection coefficient of our $4$-bits DPS-O.
    \label{fig:DPSresult}}
\end{figure}

In \eqref{eq: DPSmodel_Spar}, the two terms interpret that the (complex-valued) signal will be in \emph{constructive} or \emph{destructive} interferences. Specifically, the first term is the reflected signal corresponding to the DPS itself, and the second term is the signal passing through the DPS. To verify the transmission-line model \eqref{eq: DPSmodel_Spar}, we measure the S parameters of the DPS without an open end at different controlled states using vector network analyzers (VNA), and substitute them into \eqref{eq: DPSmodel_Spar}. The corresponding results are shown in Fig. \ref{fig:DPSresult}, which illustrates that the results based on \eqref{eq: DPSmodel_Spar} are close to those measured directly from the DPS-O through the VNA. Notably, because the reflected signal passes through the DPS twice (round trip), there is a $2 \times 360^{\circ}$ change in the phase domain. Moreover, the curve of attenuation appears in two cycles, where each corresponds to a $360^{\circ}$ change in the phase domain.

Next, we present a further modeling of the phases and attenuations at different controlled states of the DPS-O.
From the results shown in Fig. \ref{fig:DPSresult}, it can be observed that the phase and attenuation show a linear relationship with the state of DPS. This inspires us to use a similar property as in the binary weighted digital-to-analog converter to model the phases and attenuations at different states. Specifically, we model the four-bit DPS through the four orders $\gamma_{1}, \ldots, \gamma_{4}$, each providing different phase shifts and attenuations, as shown in Fig. \ref{fig:RISelement}(b).
To represent the ON/OFF state of the four orders, we use a four-bit digital input code $(b_1, b_2, b_3, b_4)$, with $1$ and $0$ representing the ON and OFF states, respectively. As a result, we can express the magnitude and phase of the reflection coefficient as the weighted-sum of the four orders, given by:
\begin{subequations} \label{eq:B2P}
\begin{align}
    |\Gamma| &= \gamma_{0} + b_1 \gamma_{1} + b_2 \gamma_{2} + b_3 \gamma_{3} + b_4 \gamma_{4} ~(\mbox{dB}), \label{eq:B2P_Amp} \\
    \measuredangle \Gamma &= \measuredangle \gamma_{0} + b_1 \measuredangle \gamma_{1} + b_2 \measuredangle \gamma_{2} + b_3 \measuredangle \gamma_{3} + b_4 \measuredangle \gamma_{4} ~(\mbox{deg}), \label{eq:B2P_Phase}
\end{align}
\end{subequations}
where $\gamma_{0}$ (dB) and $\measuredangle \gamma_{0}$ (deg) denote the attenuation and phase at the reference state $(0, 0, 0, 0)$, respectively. The weighted-sum equations enable us to convert a digital binary number into an equivalent analog output signal proportional to the values of the attenuation and phase. With \eqref{eq:B2P_Amp} and \eqref{eq:B2P_Phase}, we can compute the reflection coefficient as follows:
\begin{equation}
\Gamma = |\Gamma| e^{j \measuredangle \Gamma}.
\end{equation}
Note that $\Gamma$ is a function of $(b_1, b_2, b_3, b_4)$, which we have omitted for ease of notation.


The estimated parameters for the four-bit DPS-O obtained through the least squares (LS) estimation are presented in Table \ref{tab:TableConstrainLS}. Notably, we restrict $\gamma_{n} \leq 0, ~\forall n$ to reflect the gain loss in dB scalar. The correspond results with the estimated parameters by \eqref{eq:B2P_Amp} and \eqref{eq:B2P_Phase} are shown in
Fig. \ref{fig:DPSresult}, which match the direct measurement results. We make two interesting observations from Table \ref{tab:TableConstrainLS}. First, the estimated phase $\measuredangle\gamma_{n}$ is approximately two times phase directly measured from the DPS (i.e., phase of $S_{21}^{\rm dps}$).
This result is reasonable because the reflected signal passes through the DPS twice. The small error can be attributed to the perfect impedance matching of the internal DPS.
Second, the constructive and destructive interferences occur when the shifted phases $\measuredangle\gamma_{n}$ are around $0$ or $180$ deg, respectively, where the attenuations are $0$ and $-2.35$ dB, respectively.

\begin{table}
\begin{center}
\begin{footnotesize}
\caption{Results from our analysis of the DPS using a constrained linear LS problem. \label{tab:TableConstrainLS}}
\begin{tabular}{|l||c||c||c||c|}
\hline
$n$-th order     & ${\gamma}_{1}$  & ${\gamma}_{2}$
                & ${\gamma}_{3}$  & ${\gamma}_{4}$ \\ \hline
Attenuation (dB) & 0  & -2.35 & -1.66  & -0.57 \\ \hline
\hline
$n$-th order     & ${\measuredangle \gamma}_{1}$  & ${\measuredangle \gamma}_{2}$
                & ${\measuredangle \gamma}_{3}$  & ${\measuredangle \gamma}_{4}$ \\ \hline
Shifted phase (deg)   & -356  & -178 & -96  & -33 \\ \hline\hline

Binary code in DPS & (1,0,0,0)  & (0,1,0,0) & (0,0,1,0)  & (0,0,0,1) \\ \hline
Shifted phase by & \multirow{2}{*}{-179}  & \multirow{2}{*}{-88}
                             & \multirow{2}{*}{-45}  & \multirow{2}{*}{-23} \\
measuring DPS (deg)       &    &    &    & \\ \hline
\end{tabular}
\end{footnotesize}
\end{center}
\end{table}

In summary, the DPS-O exhibits phase-dependent attenuation in each state due to the different levels of constructive or destructive interferences, as shown in \eqref{eq: DPSmodel_Spar}.
The different orders in the DPS-O contribute different levels of attenuation loss and phase shifts, which are modeled by \eqref{eq:B2P_Amp} and \eqref{eq:B2P_Phase}.

Our proposed  analyzing methods, which include the transmission-line model presented in \eqref{eq: DPSmodel_Spar} and the constrained linear LS estimation through \eqref{eq:B2P_Amp} and \eqref{eq:B2P_Phase}, can be easily extended to other architectures of phase shifters. We present two examples to illustrate this point.
First, our DPS model in Fig. \ref{fig:RISelement}(b) is based on the multiple cascaded one-bit phase shifters \cite{Lee-2015}. When the other phase shifter is composed of multiple cascaded delay lines connected with switch components, its equivalent circuit model is similar to the DPS, and our analyzing methods can be applied.
Second, in \cite{Dunna-2020}, RF switches are used to connect four open-ended delay lines in parallel. Each open-ended delay line can be viewed as our DPS model with only one controlled state providing the fixed phase shift. Hence, our analyzing methods can be applied in this case as well.

\subsection*{B. Radar Cross Section of RIS}

Next, we clarify the relationships between the RIS as a scatterer and as a radiator. In reality, the radar cross section (RCS) of a RIS is composed of two components: antenna mode (AM) and structural mode (SM) \cite{Hansen-1989,Gan-2020}. To illustrate this characteristic, we represent the signal going towards and reflected from the feed port as $V_1^{-}$ and $V_1^{+}$, as shown in Fig. \ref{fig:RISelement}(c). Additionally, we represent the incoming and outgoing EM waves as $V^{\rm in}$ and $V^{\rm out}$. Then, a general scattering matrix relating these two is given by \cite{Hansen-1989}
\begin{equation} \label{eq: EM_Sparameters}
    \left[    \begin{array}{c}
    V_{1}^{+}       \\
    V^{\rm out}
    \end{array}    \right]
     =
    \left[    \begin{array}{cc}
    S_{00}      & S_{01}  \\
    S_{10}      & S_{11}
    \end{array}     \right]
    \left[    \begin{array}{c}
    V_{1}^{-} \\
    V^{\rm in}
    \end{array}    \right],
\end{equation}
where $S_{\cdot, \cdot}$ denotes the S-parameters of the antenna with $S_{00}$ and $S_{11}$ being the reflection coefficients of the antenna port and the metal surface, respectively.

Using a similar method as in \eqref{eq: DPSmodel_Spar}, the relationship between the incident EM waves $V^{\rm in}$ and the outgoing EM wave $V^{\rm out}$ can be formulated as
\begin{equation} \label{eq: RISmodel_Spar}
    V^{\rm out}
    =
    {S_{11} V^{\rm in}}
    +
    \frac{S_{10} \Gamma}{1 - \Gamma S_{00}}
    {S_{01} V^{\rm in}},
\end{equation}
where $\Gamma$ is the reflection coefficient of DPS-O explained in Section II-A. Because the second term in the right hand side of \eqref{eq: RISmodel_Spar} involves the effect of DPS-O, we refer it as the AM of RIS, and the first term as the SM of RIS. Specifically, the AM of RIS arises from the re-radiation of antennas under an incident plane wave, which is related to the radiation performance and loads of the antennas.

The AM of RIS can be affected by the different load ends in \eqref{eq: DPSmodel_Spar}. To better understand this, consider a scenario where the antenna is directly connected to a load end without a DPS. In this scenario, we obtain $\Gamma = \Gamma_{\rm end}$ in \eqref{eq: DPSmodel_Spar}, and there are three types of load ends to consider: open end, short end, and 50-Ohm end. For the open and short ends, we have $\Gamma_{\rm end} = -1$ and $1$, respectively. These load ends only change the AM of RIS in \eqref{eq: RISmodel_Spar}, but the antenna still retransmits the injected signal. However, with a 50-Ohm end, $\Gamma_{\rm end} = 0$, resulting in the removal of the AM of RIS and only leaving the SM of RIS. Consequently, the SM of RIS always exists while the AM of RIS is the focus because it can tailor the incident EM wave.

As the incident EM wave passes through the substrate and is affected by $S_{11}$, the controlling method for the SM of RIS can be achieved by manipulating the controllable properties of the substrate material, such as using liquid crystals (LCs) \cite{Perez-2013,Li-2021,Guirado-2022,Kim-2022}. On the other hand, when switch components of RIS \cite{Ren-2022,Pei-2021,Dunna-2020,Rao-2022} are used as the controlling method, the controllable signal is the AM of RIS, which is determined by the antenna resonant mode.

Given \eqref{eq: RISmodel_Spar}, we can identify three intuitive ways to improve the performance of the AM of RIS in practice. The first method is to reduce the size of the ground plane, which reduces the SM of RIS caused by scattering and reflecting and does not interfere with the AM of RIS. The second method is to increase the number of RIS elements on the same-sized metal plate, which maintains the SM of RIS and improves the AM of RIS through a better antenna array factor \cite{ANT-BOOK}. The third method is to use active RIS elements \cite{Zhang-2022}, which directly amplify the signal power of the AM of RIS.

\section*{III. Channel Model for RIS}

In this section, we introduce a propagation channel model for a RIS element based on the 3GPP spatial channel model (SCM), which accounts for the polarization of the RIS. The SCM is a statistical ray-tracing method that calculates the angle of departure (AoD), angle of arrival (AoA), and total path length for each path, resulting in a signal delay. The polarization of the antenna is an essential consideration when modeling wireless communication systems as it can impact the AoD and AoA of the signal. The SCM distinguishes between propagation and antenna effects. Although QuaDRiGa \cite{Jaeckel-2012,Jaeckel-2014} has previously provided a polarization model for antennas used as transmitters or receivers, it does not include the presence of an RIS in the channel model. Our proposed channel model takes into account the polarization of the RIS, making it more accurate and practical for real-world scenarios. For simplicity, we consider only two paths, one line-of-sight (LoS) path and one RIS path, but the model can be extended to multiple non-line-of-sight (NLoS) paths. The antenna model is introduced in Section III.A, and the propagation channels for the LoS and NLoS paths are formulated in Sections III.B and C, respectively. The model is validated by measurements in Section III.D.

\subsection*{A. Short Introduction to Antenna Model}

\begin{figure}
    \centering
    \resizebox{3.6in}{!}{%
    \includegraphics*{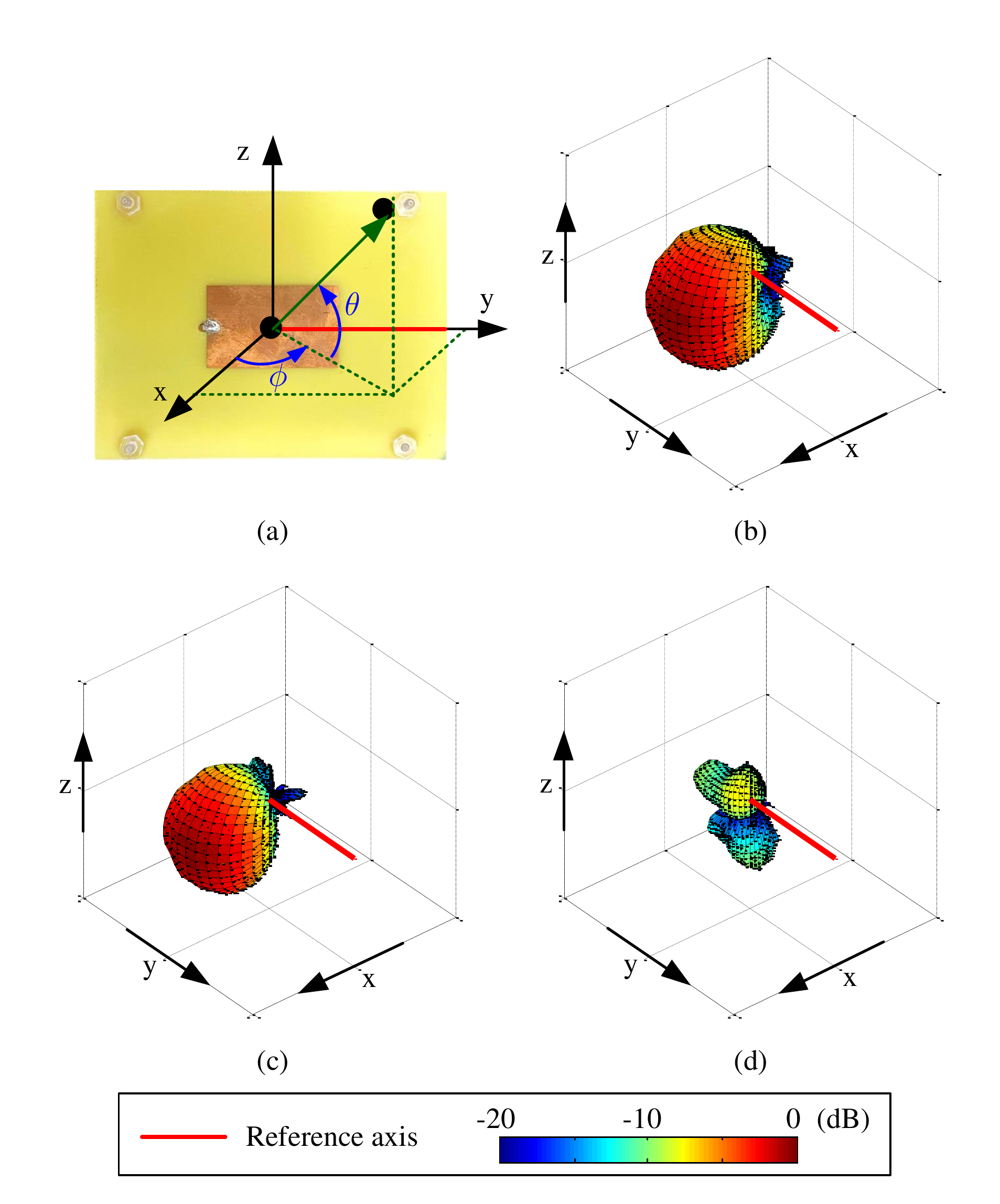} }%
    \caption{(a) Coordinate system on the patch antenna. The patch antenna's normalized (b) total antenna gain, (c) vertical antenna gain $\qE^{\rm V}$, and (d) horizontal antenna gain $\qE^{\rm H}$ for all angle pairs by being measured in the anechoic chamber.
    \label{fig:EMangle}}
\end{figure}

Assuming a planar waveform propagating along the direction $(\theta,\phi)$, where $\theta$ and $\phi$ are the elevation and azimuth angles, respectively, as illustrated in Fig. \ref{fig:EMangle}(a), we define the elevation angle ${\theta \in [\pi/2, -\pi/2]}$ as measured upward from the x-y plane and the azimuth angle ${\phi \in [-\pi, \pi]}$ as measured clockwise from the positive x-axis. The radiation pattern, also referred to as the field pattern or antenna gain, is characterized by the directional response of the antenna. Specifically, the antenna response at the angle pair $(\theta, \phi)$ consists of vertically and horizontally polarized components given by:
\begin{equation} \label{eq:AntPattern}
\qE(\theta, \phi) = {\left[ E^{\rm V}(\theta,\phi), \, E^{\rm H}(\theta,\phi) \right]}^T.
\end{equation}

The polarimetric radiation pattern is convenient to describe in a global coordinate system (GCS), and therefore all calculations can be performed in the GCS. The GCS is defined by the coordinates $(x,y,z)$ and the associated elevation and azimuth angles $(\theta, \phi)$, as shown in Fig. \ref{fig:EMangle}(a). When the orientation of the antenna changes, the radiation pattern can be transformed to maintain alignment with the GCS, making it easier to read the antenna's polarimetric responses using the departure and arrival angles of propagation paths. We will provide a detailed description of the transformation process. Due to the principle of reciprocity, the radiation pattern and polarization of a transmit antenna are the same as those of a receive antenna \cite{ANT-BOOK}. Therefore, we will only discuss the radiation pattern and polarization of a transmitting antenna in detail.

When the orientation of the antenna changes, two effects must be considered: the radiation pattern and the rotation of polarization in the polar-spherical basis. To proceed, we denote $(\tilde{x},\tilde{y},\tilde{z})$ as the local coordinate system of the rotated antenna, with associated elevation and azimuth angles $(\tilde{\theta}, \tilde{\phi})$. Moreover, to describe the antenna's rotation, we define the rotation matrix \eqref{eq:RotMatrix} at the top of the next page,
\begin{figure*}
\begin{equation}
\label{eq:RotMatrix}
    \qR =
    \left[    \begin{array}{ccc}
    \cos(r_z)\cos(r_y) &
    \cos(r_z)\sin(r_y)\sin(r_x) - \sin(r_z)\cos(r_x) &
    \cos(r_z)\sin(r_y)\cos(r_x) + \sin(r_z)\sin(r_x) \\
    \sin(r_z)\cos(r_y) &
    \sin(r_z)\sin(r_y)\sin(r_x) + \cos(r_z)\cos(r_x) &
    \sin(r_z)\sin(r_y)\cos(r_x) - \cos(r_z)\sin(r_x) \\
    -\sin(r_y)          &
    \cos(r_y)\sin(r_x)           &
    \cos(r_y)\cos(r_x)
    \end{array}     \right]
\end{equation}
\hrule
\end{figure*}
where $r_x$, $r_y$, and $r_z$ represent the orientation angles of the antenna rotated along the x-axis, y-axis, and z-axis, respectively.

We first consider the rotated coefficients of the radiation pattern. Assuming that before antenna rotation, the original transmitted signal direction, i.e., AoD, is in the angle pair $(\theta, \phi)$. Recall that the radiation pattern at the angle pair is given by \eqref{eq:AntPattern}. When the antenna's orientation changes, the radiation pattern has to be read at the different AoD $(\tilde{\theta}, \tilde{\phi})$. Because the rotation matrix \eqref{eq:RotMatrix} is defined in Cartesian coordinates, the signal direction should be transformed from spherical coordinates into Cartesian coordinates. The angle pair in spherical coordinates $(\theta, \phi)$ can be transformed into Cartesian coordinates by
\begin{equation} \label{eq:OrgRadVec}
    \qs(\theta, \phi)
     =
    {\left[
    \cos(\theta)\cos(\phi),\,
    \cos(\theta)\sin(\phi),\,
    \sin(\theta) \right]^T}.
\end{equation}
With the antenna rotation expressed by the rotation matrix $\qR$, the signal direction is translated to
\begin{equation}
  [s_{\tilde{x}}, s_{\tilde{y}}, s_{\tilde{z}}]^T =\qR^{T} \qs .
\end{equation}
Now, the AoD of the rotated antenna can be calculated by
\begin{equation}
 \tilde{\theta} = \sin^{-1}( s_{\tilde{z}}), ~~ \tilde{\phi} = \tan^{-1}{\left(\frac{s_{\tilde{y}}}{s_{\tilde{x}}}\right)}.
\end{equation}
Finally, the coefficients of the rotated pattern are obtained by reading the original pattern $\qE$ at the rotated AoD $(\tilde{\theta}, \tilde{\phi})$ as
\begin{equation}
 \qE(\tilde{\theta}, \tilde{\phi}) = {\left[E^{\rm V}(\tilde{\theta}, \tilde{\phi}), \, E^{\rm H}(\tilde{\theta}, \tilde{\phi})\right]^T}.
\end{equation}

Second, we consider the rotation of polarization. The radiation pattern is defined in spherical coordinates. Therefore, we define a transformation matrix from spherical coordinates in the direction $(\theta, \phi)$ to Cartesian coordinates as follows:
\begin{equation} \label{eq:OrgPolar}
    \qT(\theta, \phi)
     =
    \left[    \begin{array}{cc}
    \sin(\theta)\cos(\phi) & -\sin(\phi) \\
    \sin(\theta)\sin(\phi) & \cos(\phi)  \\
    -\cos(\theta)           & 0
    \end{array}     \right],
\end{equation}
where the first and second columns are the transformations of the vertical and horizontal polarizations into Cartesian coordinates, respectively. Since (\ref{eq:OrgPolar}) is an orthogonal matrix, its inverse is its transpose. Using the transformation matrix, we can convert the coefficients of the rotated pattern from spherical coordinates to Cartesian coordinates using the following equation:
\begin{equation}  \label{eq:E_local}
     [E_{\tilde{x}}, E_{\tilde{y}}, E_{\tilde{z}}]^T = \qT(\tilde{\theta}, \tilde{\phi})  \qE(\tilde{\theta}, \tilde{\phi}).
\end{equation}
Note that (\ref{eq:E_local}) is in the local Cartesian coordinate system of the rotated antenna. It can be mapped to the global Cartesian coordinate system by multiplying the rotation matrix, yielding:
\begin{equation}  \label{eq:E_global}
     [E_{x}, E_{y}, E_{z}]^T = \qR  \qT(\tilde{\theta}, \tilde{\phi})  \qE(\tilde{\theta}, \tilde{\phi}).
\end{equation}
Finally, we transform the global Cartesian coordinates to the global spherical coordinates. Combining everything, the rotated polarization is obtained as follows:
\begin{equation} \label{eq:RotMat}
    \sfqE(\theta, \phi) = \underbrace{\qT(\theta, \phi)^T  \qR   \qT(\tilde{\theta}, \tilde{\phi})}_{\triangleq \qM_{\rm o} } \cdot \qE(\tilde{\theta}, \tilde{\phi}).
\end{equation}
Here, ${\sfqE(\theta, \phi) = [\sfE^{\rm V}(\theta, \phi), , \sfE^{\rm H}(\theta, \phi)]^T}$ is the final radiation pattern of the vertical and horizontal polarizations after rotation in the global angle pair $(\theta, \phi)$. In (\ref{eq:RotMat}), ${\qM_{\rm o} = \qT(\theta, \phi)^T \qR \qT(\tilde{\theta}, \tilde{\phi}) }$ is the $2 \times 2$ polarization rotation matrix that captures the antenna's orientation change.

\subsection*{B. Model for LoS Path}

In this subsection, we describe the channel model for a LoS scenario with a transmit antenna and a receive antenna. To ensure accurate description of the radiation patterns in the same baseline, all the angles must be in the GCS. The transmit antenna and receive antenna are expressed as $\sfqE^{\rm t}(\theta^{\rm t}, \phi^{\rm t})$ and $\sfqE^{\rm r}(\theta^{\rm r}, \phi^{\rm r})$, where $(\theta^{\rm t}, \phi^{\rm t})$ and $(\theta^{\rm r}, \phi^{\rm r})$ represent the AoD and AoA, respectively.

Since the polarization direction of the transmit antenna and the receive antenna is observed in opposite reference points, a transformation of polarization direction is necessary, which is expressed as
\begin{equation}
    \qM_{\rm d}
     =
    \left[    \begin{array}{cc}
    1 & 0             \\
    0 & -1
    \end{array}     \right].
\end{equation}
Therefore, the LoS channel coefficient, which takes into account the orientation of the antennas, is formulated as
\begin{equation} \label{eq:LOS}
    C^{{\rm t}, {\rm r}}
     =
    \sqrt{P^{\rm t} L^{{\rm t}, {\rm r}}} \sfqE^{\rm r}(\theta^{\rm r}, \phi^{\rm r})^T \qM_{\rm d}
    \sfqE^{\rm t}(\theta^{\rm t}, \phi^{\rm t}) e^{-j (\psi^{\rm t} + \psi^{\rm r} + k d^{{\rm t}, {\rm r}})},
\end{equation}
where $P^{\rm t}$ is the transmitted power, $L^{{\rm t}, {\rm r}} = (\lambda / 4 \pi d^{{\rm t}, {\rm r}})^2$ represents the path loss, $d^{{\rm t}, {\rm r}}$ is the distance between the transmit antenna and the receive antenna, $\lambda$ is the frequency of the EM wave, $k = 2 \pi / \lambda$ is the wave number, and $\psi^{\rm t}$ and $\psi^{\rm r}$ are the initial phases of the transmit antenna and receive antenna respectively, caused by the manufacturing process.

In \eqref{eq:LOS}, the polarization of antennas is considered.
Furthermore, the factor $|\sfqE^{\rm r}(\theta^{\rm r}, \phi^{\rm r})^T \qM_{\rm d} \sfqE^{\rm t}(\theta^{\rm t}, \phi^{\rm t})|$ reflects the mismatch between the polarizations of the transmit and receive antennas. The definition of the polarization efficiency (or polarization mismatch factor) can be found in \cite{ANT-BOOK}.

\subsection*{C. Model for RIS Path}

Next, we extend (\ref{eq:LOS}) to describe the channel model for a RIS. For the sake of simplicity, we consider the case where there is only one incident wave. To aid understanding, we define some useful notations. Let $(\theta^{{\rm x},{\rm y}}, \phi^{{\rm x},{\rm y}})$ denote the angle pair of the vector pointing from ${\rm x}$ to ${\rm y}$, where ${\rm x}, {\rm y} \in \{ {\rm t}, {\rm r}, {\rm ris} \}$. Similarly, let $d^{{\rm x},{\rm y}}$ represent the distance between ${\rm x}$ and ${\rm y}$. It is worth noting that the order of the superscripts $^{\rm x}$ and $^{\rm y}$ affects the angle, but not the distance.

In the case of the AM of RIS, the signal received by the RIS is affected by the polarization efficiency. To account for this, we define the polarimetric response of the RIS as ${\sfqE^{\rm ris}(\theta, \phi) \in \bbC^{2 \times 1}}$. The signal then passes through a DPS and is reflected by an open end. The phase-dependent attenuation of the reflection coefficient $\Gamma$, caused by constructive or destructive interferences of DPS-O, must be considered. Finally, the signal is delivered to the receiver.

Using a similar concept as in \eqref{eq:LOS}, the channel coefficient of the AM of RIS can be formulated as follows:
\begin{equation} \label{eq:RISinj}
\begin{aligned}
    C_{\rm am}^{{\rm ris}}
    &=  \sqrt{P^{\rm t} L^{{\rm ris}, {\rm r}} L^{{\rm t}, {\rm ris}}} \\
    &~~~ \cdot \underbrace{{\sfqE^{\rm r}(\theta^{{\rm r},{\rm ris}}, \phi^{{\rm r},{\rm ris}})}^T \qM_{\rm d}  \sfqE^{\rm ris}(\theta^{{\rm ris},{\rm r}}, \phi^{{\rm ris},{\rm r}})}_{{\rm ris}\rightarrow {\rm r}}   \\
    &~~~ \cdot \Gamma \cdot \underbrace{{\sfqE^{\rm ris}(\theta^{{\rm ris},{\rm t}}, \phi^{{\rm ris},{\rm t}})}^T  \qM_{\rm d} \sfqE^{\rm t}(\theta^{{\rm t},{\rm ris}}, \phi^{{\rm t},{\rm ris}})}_{{\rm t}\rightarrow {\rm ris}}  \\
    &~~~  \cdot e^{- j [\psi^{\rm t} + \psi^{\rm r} + \psi^{\rm ris} + k (d^{{\rm ris}, {\rm r}}+d^{{\rm t}, {\rm ris}})]},
\end{aligned}
\end{equation}
where $\psi^{\rm ris}$ is the initial phase from the antenna of RIS element, $L^{{\rm t}, {\rm ris}}=(\lambda / 4 \pi d^{{\rm t}, {\rm ris}})^2$ and $L^{{\rm ris}, {\rm r}}=(\lambda / 4 \pi d^{{\rm ris}, {\rm r}})^2$ are the path loss from the transmit antenna to the RIS, and from the RIS to the receive antenna, respectively. In \eqref{eq:RISinj}, the transformation of polarization corresponding to the AM of RIS can be captured by
\begin{equation} \label{eq:INJ}
    \qM_{\rm am}
     = \Gamma \cdot \sfqE^{\rm ris}(\theta^{{\rm ris},{\rm r}}, \phi^{{\rm ris},{\rm r}}) {\sfqE^{\rm ris}(\theta^{{\rm ris},{\rm t}}, \phi^{{\rm ris},{\rm t}})}^T.
\end{equation}
By using the above expression, we can be expressed \eqref{eq:RISinj} as
\begin{equation} \label{eq:RISinj_wM}
\begin{aligned}
    C_{\rm am}^{{\rm ris}}
    &=  \sqrt{P^{\rm t} L^{{\rm ris}, {\rm r}} L^{{\rm t}, {\rm ris}}} \\
    &~~~ \cdot {\sfqE^{\rm r}(\theta^{{\rm r},{\rm ris}}, \phi^{{\rm r},{\rm ris}})}^T \qM_{\rm d} \qM_{\rm am} \qM_{\rm d} \sfqE^{\rm t}(\theta^{{\rm t},{\rm ris}}, \phi^{{\rm t},{\rm ris}})  \\
    &~~~  \cdot e^{- j [\psi^{\rm t} + \psi^{\rm r} + \psi^{\rm ris} + k (d^{{\rm ris}, {\rm r}}+d^{{\rm t}, {\rm ris}})]} .
\end{aligned}
\end{equation}

To model the SM of RIS, we introduce the transformation of polarization for the scattered and reflected waves, respectively \cite{AEE-BOOK}:
\begin{equation} \label{eq:ScatterReflect}
    \qM_{\rm sca}
    =
    \left[    \begin{array}{cc}
    S^{\rm VV} & S^{\rm VH}            \\
    S^{\rm HV} & S^{\rm HH}
    \end{array}     \right],~~
    \qM_{\rm ref}
    =
    \left[    \begin{array}{cc}
    R^{\perp}   & 0             \\
    0           & R^{\parallel}
    \end{array}     \right],
\end{equation}
where the parameters are functions of the plate size and angles of the incident, scattered, and reflected waves. We do not provide the detailed expressions of these parameters, but readers can refer to \cite{AEE-BOOK} for more information.

Because of the approximate shape, approximate material parameters, and neglect of edge effects assumed in \eqref{eq:ScatterReflect}, some tuning coefficients are necessary. Thus, the transformation of polarization corresponding to the SM of RIS can be expressed as
\begin{equation} \label{eq:SM}
    \qM_{\rm sm}
     = c_{\rm s}\qM_{\rm sca} + c_{\rm r}\qM_{\rm ref},
\end{equation}
where $c_{\rm s}$ and $c_{\rm r}$ are the tuning coefficients for the scattered and reflected fields, respectively.

The channel coefficient corresponding to the RIS's metal material is given by:
\begin{equation} \label{eq:RISmet}
\begin{aligned}
    C_{\rm sm}^{{\rm ris}}
    =
    & \sqrt{P^{\rm t} L^{{\rm ris}, {\rm r}} L^{{\rm t}, {\rm ris}}} \\
    & \cdot {\sfqE^{\rm r}(\theta^{{\rm r},{\rm ris}}, \phi^{{\rm r},{\rm ris}})}^T \qM_{\rm d}  \qM_{\rm sm} \qM_{\rm d} \sfqE^{\rm t}(\theta^{{\rm t},{\rm ris}}, \phi^{{\rm t},{\rm ris}}) \\
    & \cdot e^{- j [\psi^{\rm t} + \psi^{\rm r} + k (d^{{\rm ris}, {\rm r}}+d^{{\rm t}, {\rm ris}})]}.
\end{aligned}
\end{equation}
It is important to note that the influence of DPS-O with $\Gamma$ in \eqref{eq:INJ} can control $C_{\rm am}^{{\rm ris}}$ in \eqref{eq:RISinj_wM}, but not $C_{\rm sm}^{{\rm ris}}$ in \eqref{eq:RISmet}.
Most literature models the signal contributed from the RIS by multiplying the reflection coefficient $\Gamma$ by $C_{\rm sm}^{{\rm ris}}$. However, this model is incorrect as it ignores the EM field generated by the antenna, i.e., the distributions of currents in the antenna.

Combining \eqref{eq:LOS}, \eqref{eq:RISinj_wM}, and \eqref{eq:RISmet}, the overall channel model of a RIS element in the LoS case is expressed as:
\begin{equation} \label{eq:All}
    C_{\rm element} = C^{{\rm t},{\rm r}} + C_{\rm am}^{{\rm ris}} + C_{\rm sm}^{{\rm ris}},
\end{equation}
where the first term is contributed by the LoS component, and the remains are contributed by the RIS. The antenna response can be directly determined by measuring in an anechoic chamber, a process commonly referred to as antenna calibration.
Notably, since the phase and amplitude variations over the antenna's diagonal are not ignored in the antenna calibration, the applicable distance between the transmitter, receiver, and RIS for this channel model only needs to be larger than the Fraunhofer distance of a single antenna, rather than the Fraunhofer array distance of a complete antenna array \cite{NF-2021}. Therefore, the proposed channel model is also suitable for the near-field beamforming of an array.

\subsection*{D. Validation}

\begin{figure}
    \centering
    \resizebox{3.4in}{!}{%
    \includegraphics*{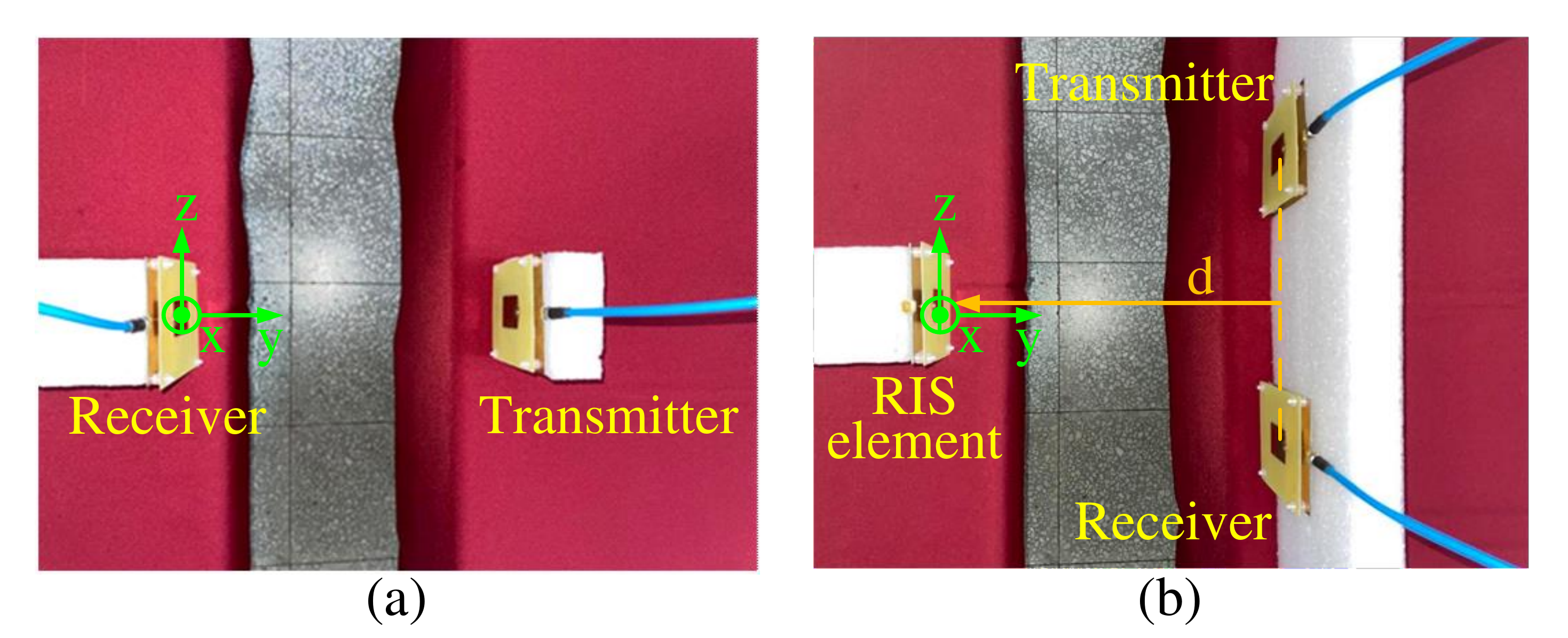} }%
    \caption{Experimental setup for the transmit antenna and the receive antenna scenario (a) without or (b) with the RIS element.
    \label{fig:Environment}}
\end{figure}

To verify the transformation of polarization in the proposed channel model, we conducted a series of experiments and simulations. The experimental scenarios are shown in Fig. \ref{fig:Environment}. Using a VNA, we measured the gain loss, defined as the received power divided by the transmitted power. The original radiation patterns of the vertical and horizontal polarizations, $\qE^{\rm V}$ and $\qE^{\rm H}$, respectively, were measured in an anechoic chamber and are shown in Figs. \ref{fig:EMangle}(c) and (d). Since the vertical and horizontal polarizations are described by the antenna gain and efficiency, we do not need to include them in \eqref{eq:LOS}, \eqref{eq:RISinj}, and \eqref{eq:RISmet}.
For the setup of our patch antenna in Fig. \ref{fig:EMangle}(a), the dominant received signal quality is the vertical polarization. Hence, by examining the orientation of the antennas with respect to the reference axis in Fig. \ref{fig:EMangle}, we can roughly determine whether the polarizations among antennas are matched or not.

\begin{figure}
    \centering
    \resizebox{3.6in}{!}{%
    \includegraphics*{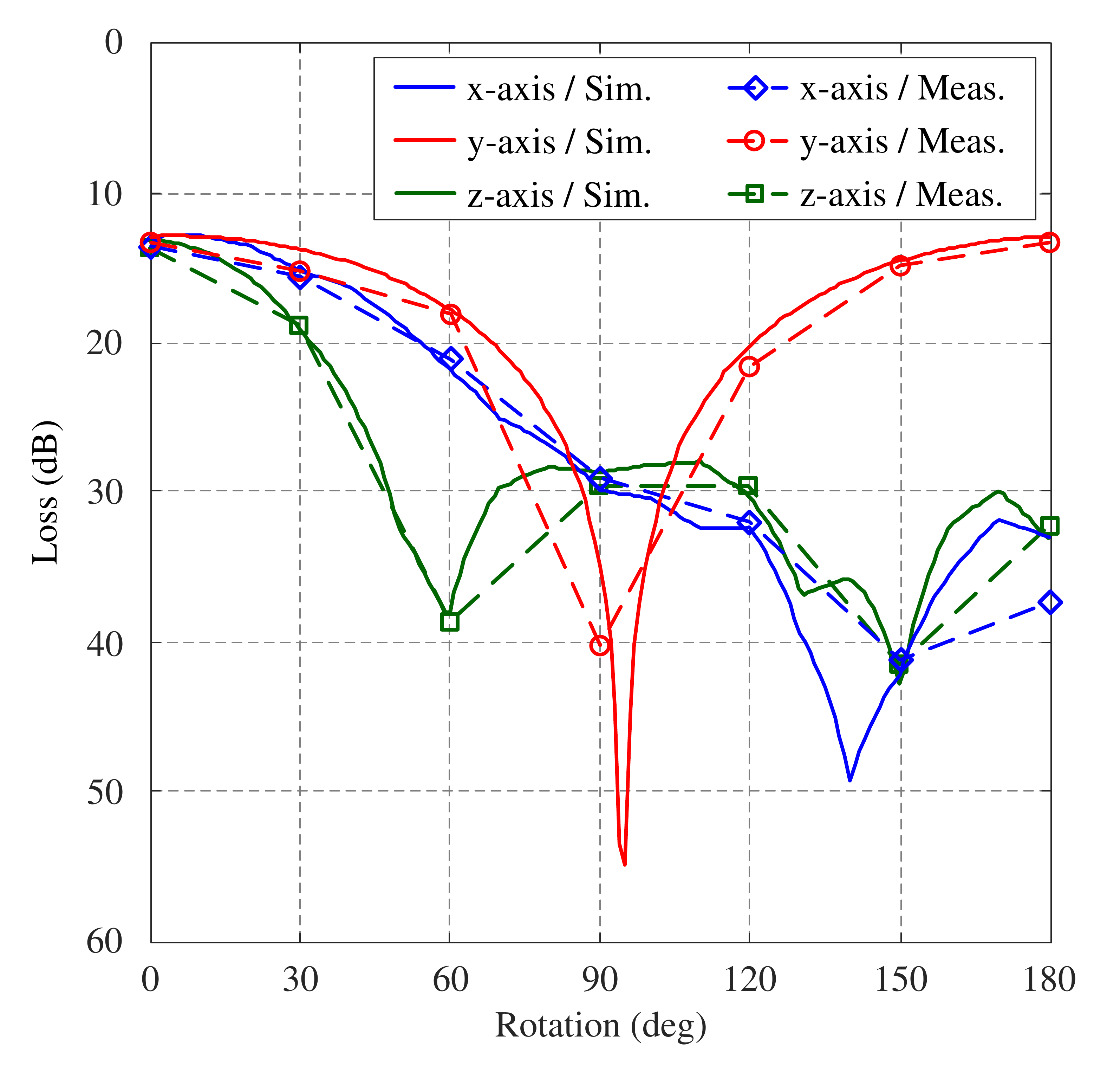} }%
    \caption{Rotation of the receive antenna corresponding to Fig. \ref{fig:Environment}(a).
    \label{fig:TxRx}}
\end{figure}

We first verified the transformation of polarization in the proposed channel model \eqref{eq:LOS} with the transmit antenna and the receive antenna, as shown in Fig. \ref{fig:Environment}(a). The receive antenna was rotated along the x-axis, y-axis, or z-axis. The simulated results of the gain loss shown in Fig. \ref{fig:TxRx} closely matched the measurement results, thereby validating the accuracy of the channel model (\ref{eq:LOS}). The gain loss changed with the rotation of the antenna. When the antenna was rotated by $90$ degrees along the y-axis, the gain loss was large due to the polarization mismatch between the transmit and receive antennas. These results illustrate the importance of considering the polarization efficiency among antennas.

\begin{figure}
    \centering
    \resizebox{3.6in}{!}{%
    \includegraphics*{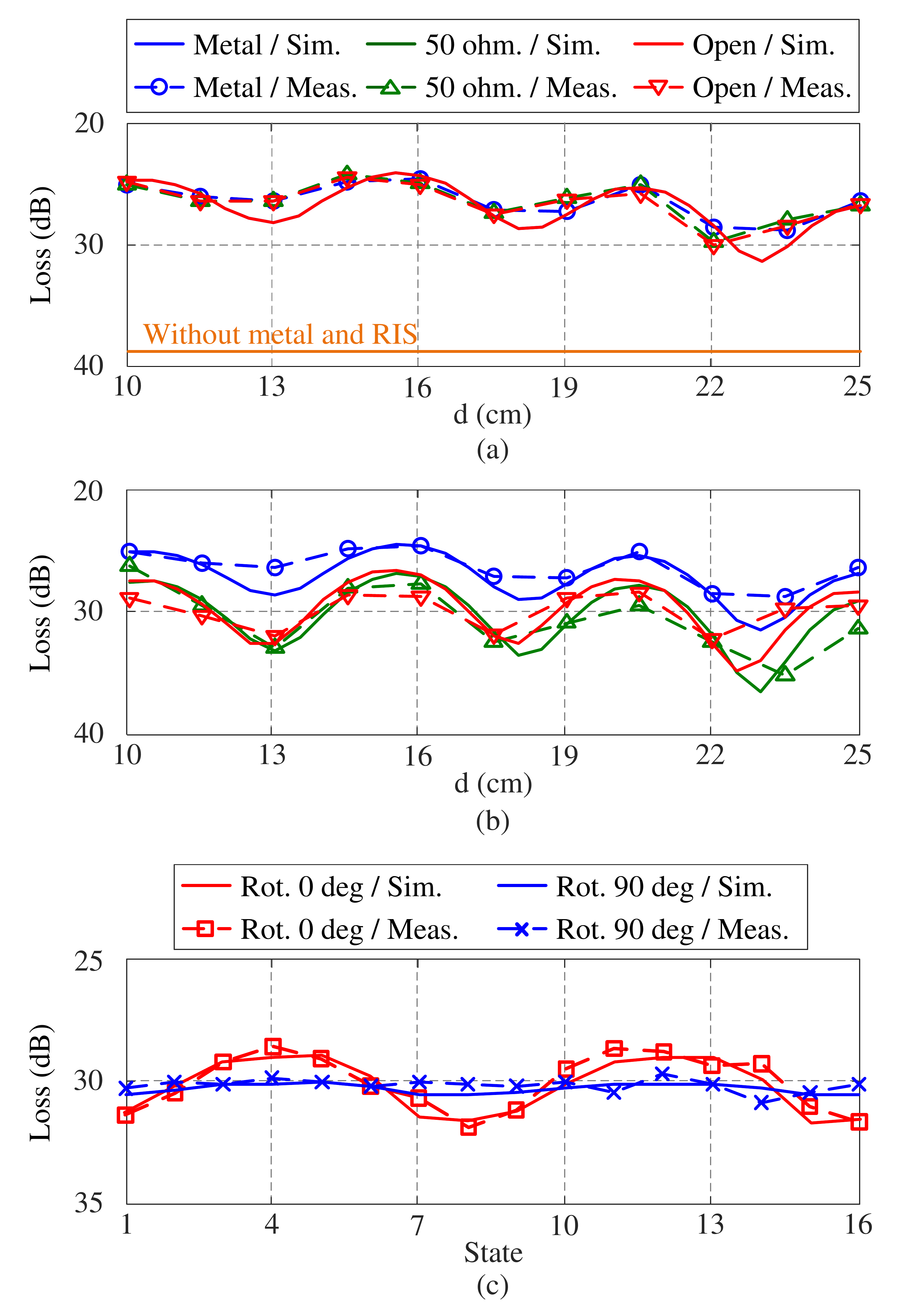} }%
    \caption{Location of the metal plate or the RIS element corresponding to Fig. \ref{fig:Environment}(b) when the (a) back or (b) front of subject fasces to the middle of the transmit antenna and the receive antenna. For the different polarization direction, (c) RIS element with a DPS replaces the RIS element in Fig. \ref{fig:Environment}(b).
    \label{fig:Tx2RIS2Rx}}
\end{figure}

In the previous experiment, we verified the LoS component in the channel model. Next, we tested the remaining two terms in the channel model \eqref{eq:All}, which are contributed by the RIS. In this experiment, we positioned the transmitter and receiver to face the same direction in order to reduce the LoS component $C^{{\rm t},{\rm r}}$ from \eqref{eq:All}. The scenario is shown in Fig. \ref{fig:Environment}(b). To simplify the analysis, we directly used an open end or 50-Ohm end on the antenna without connecting to a DPS. We analyzed different distances between the RIS element and the middle location of the transmit and receive antennas. We tested three cases, including a same-sized metal plate, an RIS with a 50-Ohm end, and an RIS with an open end. The results are shown in Fig. \ref{fig:Tx2RIS2Rx}(a) when the back of the antenna faces the middle location, i.e., the rotation of $180$ degrees along the x-axis.

On the one hand, because the RIS cannot radiate the injected signal with a 50-Ohm end, whose $\Gamma_{\rm end} = 0$, its channel model is the same as that of the same-sized metal plate, and the channel coefficient $C_{\rm am}^{{\rm ris}} = 0$ in \eqref{eq:All}. On the other hand, because the radiation pattern at the back of the antenna is weak, the channel coefficient $C_{\rm am}^{{\rm ris}}$ in \eqref{eq:All} is small. As a result, the losses of the three cases are almost the same. The simulated results based on \eqref{eq:All} closely matched the measurement results.

Furthermore, we analyzed the scenario when the front of the antenna faces the middle location, and the results are shown in Fig. \ref{fig:Tx2RIS2Rx}(b). The patch antenna is comprised of two metal plates, a radiating metal pattern at the first layer, and a ground plane at the second layer, with the size of the radiating metal pattern based on the operating band. Some incident waves can pass through the lossy FR-4 material of the front layer to the integral metal plate of the second layer, resulting in weaker energy of the passing waves. The SM of RIS is determined by the incident waves and the weak passing waves, resulting in weaker energy than those directly used by the same-sized metal plate. Therefore, the loss of the RIS with a 50-Ohm end is larger than that of the same-sized metal plate, which verifies the existence of $C_{\rm sm}^{{\rm ris}}$. When the distance increases, the losses of the RIS with open end and 50-Ohm end diverge, as the propagation angle of the AM of RIS becomes closer to the main lobe of the patch antenna. This observation indicates that the additional signal is contributed from $C_{\rm am}^{{\rm ris}}$.

Next, we replaced the RIS element at $d = 25$ in Fig. \ref{fig:Environment}(b) with the RIS element with a DPS-O in Fig. \ref{fig:RISelement}. We considered all DPS states when the RIS element was rotated by $0$ or $90$ degrees along the y-axis, and the results are shown in Fig. \ref{fig:Tx2RIS2Rx}(c).\footnote{The source code of Fig. \ref{fig:Tx2RIS2Rx}(c) is available on GitHub: \url{https://github.com/icefreeman123/Matlab_RIS_ChannelModel}.}
When the angle of rotation is $0$ degrees, the polarization among the transmit, receive, and RIS antennas is matched. Therefore, different DPS states can control the received signal in constructive or destructive interference. However, when the angle of rotation is $90$ degrees, the polarization among antennas is mismatched by the RIS antenna, and the phenomenon of interference does not occur. These observations confirm the necessity of considering the polarization efficiency in the RIS scenario, and verify the model of DPS-O in Section II-A and the channel model that considers the polarization of RIS in Section II-E.

\begin{figure}
    \centering
    \resizebox{3.6in}{!}{%
    \includegraphics*{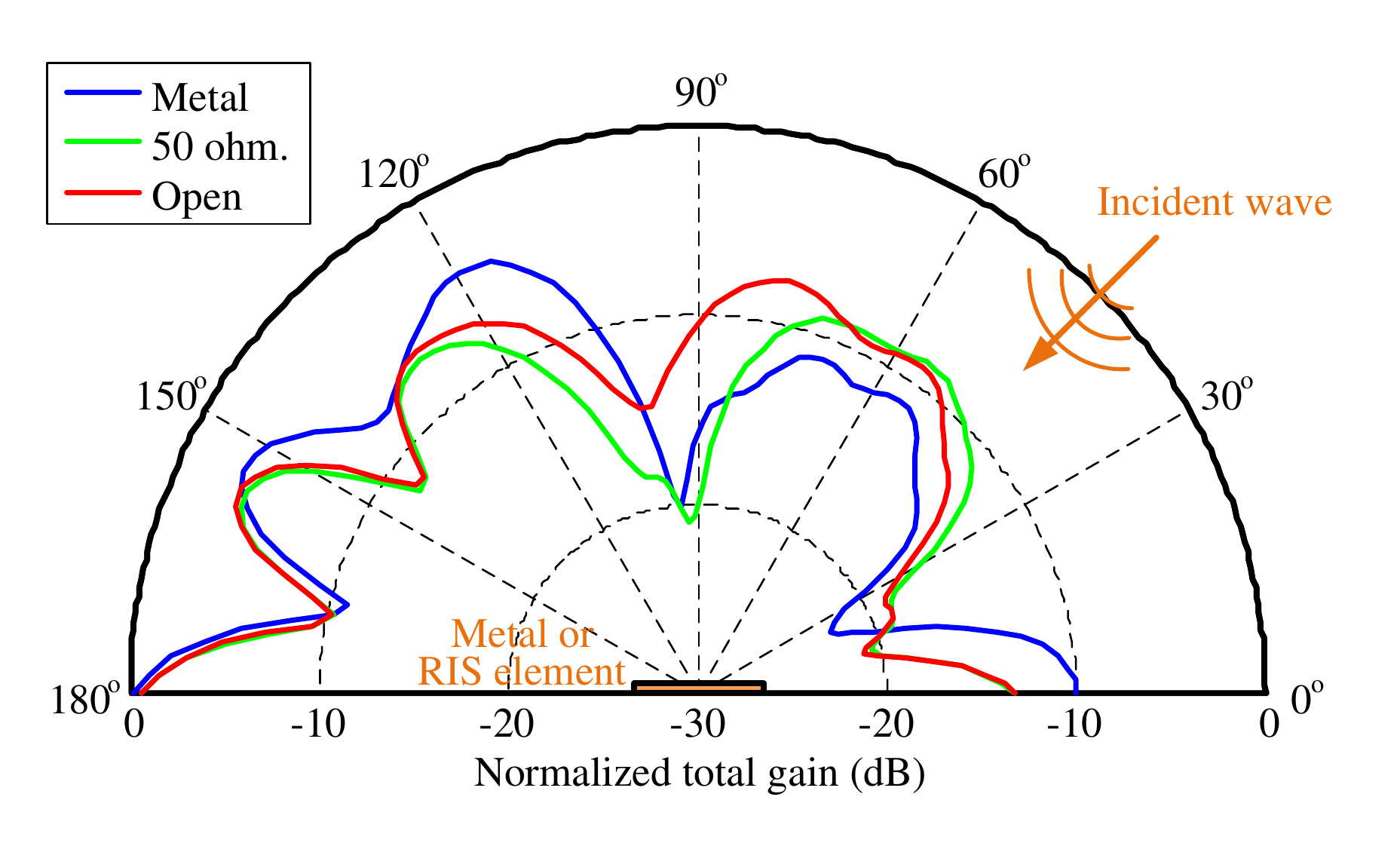} }%
    \caption{Simulated radiation patterns of the metal plate or RIS element.
    \label{fig:FEBI}}
\end{figure}

To verify the proposed channel model of the RIS element, we used ANSYS HFSS finite element boundary integral (FE-BI) to plot radiation patterns as shown in Fig. \ref{fig:FEBI}. A patch antenna transmitting EM waves at a 45-degree angle was used as the transmit antenna, and a metal plate or RIS element of the same size was placed at the center of a half circle. The simulated radiation pattern is the power measured at different locations around the circle with an angle from 0 to 180 degrees. The results show that the radiation patterns of the metal plate and RIS element with a 50-Ohm end are similar, with the reflection angle at about 135 degrees and scattered fields at other angles except for the reflection angle. This is because the RIS element cannot retransmit the injection signal with a 50-Ohm end. On the other hand, comparing the RIS element with an open end to one with a 50-Ohm end, it is found that the difference between the two patterns is only at around 90 degrees, which is the main lobe of the antenna. This result verifies that the signal is injected into the antenna and retransmitted by the RIS.

\section*{IV. Blind Controlling Algorithm}

After developing a channel model for a single RIS element, we extend it to a RIS array. The general design of antennas in the array is such that they are placed at a sufficient distance from each other to avoid coupling, hence we assume that the mutual coupling effect is negligible. In a LoS scenario, the channel model of a RIS array is the sum of all $N_{\text{ris}}$ RIS elements and can be expressed as follows:
\begin{equation} \label{eq:AllArray}
    C_{\rm array} = C^{{\rm t},{\rm r}} + \sum_{n=1}^{N_{\rm ris}} {\left(C_{{\rm am},n}^{{\rm ris}} + C_{{\rm sm},n}^{{\rm ris}}\right)},
\end{equation}
where $C_{{\text{am}},n}^{{\text{ris}}}$ is the AM of the $n$-th RIS element, and $C_{{\text{sm}},n}^{{\text{ris}}}$ is the AM of the $n$-th RIS element. The initial phases of the RIS elements may not be identical due to the manufacturing process. However, these phases can be treated as additional fixed phases that need to be compensated by the DPSs. Although the same combination of DPS states may result in different $C_{\rm array}$ in simulation and measurement, their tendencies are similar. Therefore, by using an exhaustive searching method for all combinations of DPS states, we can explain why the channel model of a RIS array \eqref{eq:AllArray} is correct, as discussed in Section V-B.

To gain a better understanding of the properties of the RIS channel model in \eqref{eq:AllArray}, we propose a controlling algorithm for all DPS states. As a controlling algorithm that relies on accurate channel information has a lot of modifying requirements and extra costs \cite{Ren-2022}, a blind method without channel information is important.

\begin{algorithm} \label{alg:BGA} \small
\caption{Blind Greedy Algorithm}
    {\bf Input:} $\qB_{\rm codebook}$, $\qB_{bit}$, $N_{\rm ris}$, $N_{\rm bit}$, $T_{\rm r}$, $T_{\rm g}$.\\
    {\bf Initialize:} $\qB_{\rm r}\leftarrow \mathbf{0}_{N_{\rm ris} \times N_{\rm bit}}$, $\qB_{\rm tmp} \leftarrow \mathbf{0}_{N_{\rm ris} \times N_{\rm bit}}$.\\
    \textbf{Random-Max Sampling:} \\
    \hspace*{0.02in} (1) Get received signal quality $P$ with $\qB_{\rm tmp}$. \\
    \hspace*{0.02in} \For{$t_{\rm r} = 1,\ldots, T_{\rm r}$ }{
        (2) Generate $\qB_{\rm tmp}$ drawn uniformly from $\qB_{\rm codebook}$. \\
        (3) Get received signal quality $P_{\rm tmp}$ with $\qB_{\rm tmp}$. \\
        (4) \If{$P_{\rm tmp} > P$}{$P \leftarrow P_{\rm tmp}$, $\qB_{\rm r} \leftarrow \qB_{\rm tmp}$.}
    }
    $(5)$ $\qB_{\rm g} \leftarrow \qB_{\rm r}$. \\
    \textbf{Greedy Searching:} \\
    \hspace*{0.02in} \For{$t_{\rm g} = 1,\ldots, T_{\rm g}$ }{
    \hspace*{0.02in} \For{$n_{\rm ris} = 1,\ldots, N_{\rm ris}$ }{
        (6) $\qB_{\rm tmp} \leftarrow \qB_{\rm g}$. \\
    \hspace*{0.02in} \For{$m_{\rm bit} = 1,\ldots, 2^{N_{\rm bit}}$ }{
        (7) $n_{\rm ris}$-th row of $\qB_{\rm tmp}$ \\
            \hspace*{0.2in} $\leftarrow$ $m_{\rm bit}$-th row of $\qB_{bit}$. \\
        (8) Get received signal quality $P_{\rm tmp}$ with $\qB_{\rm tmp}$. \\
        (9) \If{$P_{\rm tmp} > P$}{$P \leftarrow P_{\rm tmp}$, $\qB_{\rm g} \leftarrow \qB_{\rm tmp}$.}
    }}}
    $(10)$ $\qB \leftarrow \qB_{\rm g}$. \\
    {\bf Output:} $\qB$.
\end{algorithm}

We propose a blind algorithm named the Blind Greedy (BG) algorithm, based on the greedy method. The algorithm aims to find the optimal combination of all DPS states, denoted as $\qB = [\qb_1, \cdots, \qb_{N_{\text{ris}}}]^T$, where each $\qb_{n_{\text{ris}}}$ is the $N_{\text{bit}}$-bit digital input code of the $n_{\text{ris}}$-th DPS. The BG algorithm consists of two main steps: Random-Max Sampling (RMS) \cite{Ren-2022} and Greedy Searching (GS).
RMS draws uniformly from an exhaustive codebook $\qB_{\text{codebook}}$ to obtain a better initial start $\qB_{\text{r}}$ for the greedy method. After $T_{\text{r}}$ iterations, the best $\qB_{\text{r}}$ in RMS is set as the initial start $\qB_{\text{g}}$ for GS. GS draws progressively from exhaustive one DPS's states, denoted as $\qB_{\text{bit}}$, and replaces the state of the $n_{\text{ris}}$-th DPS using the greedy method. After $T_{\text{g}}$ iterations, the best $\qB_{\text{g}}$ in GS is set as the final optimal combination $\qB$.

Thanks to the uniform sampling over the entire searching space, RMS has the ability to avoid local optimal combinations. However, during RMS, the received signal quality may change drastically. In contrast, the received signal quality of GS changes smoothly and its tendency is to improve. Therefore, the BG algorithm combines a few iterations of RMS and GS to achieve a rough and refined optimal combination. The outline of the BG algorithm is presented in Algorithm~\ref{alg:BGA}.

\section*{V. Experiments, Simulations, and Discussions}

In \eqref{eq:AllArray}, we propose a new channel model that takes into account the polarization of a RIS. The received signals from the RIS elements are separated into two types: $C_{{\text{am}},n}^{{\text{ris}}}$ and $C_{{\text{sm}},n}^{{\text{ris}}}$. In order to analyze the properties of this model, we conduct a series of experiments and simulations using the BG algorithm, which are described in the following subsections.

\subsection*{A. Experiments}

\begin{figure}
    \centering
    \resizebox{3.6in}{!}{%
    \includegraphics*{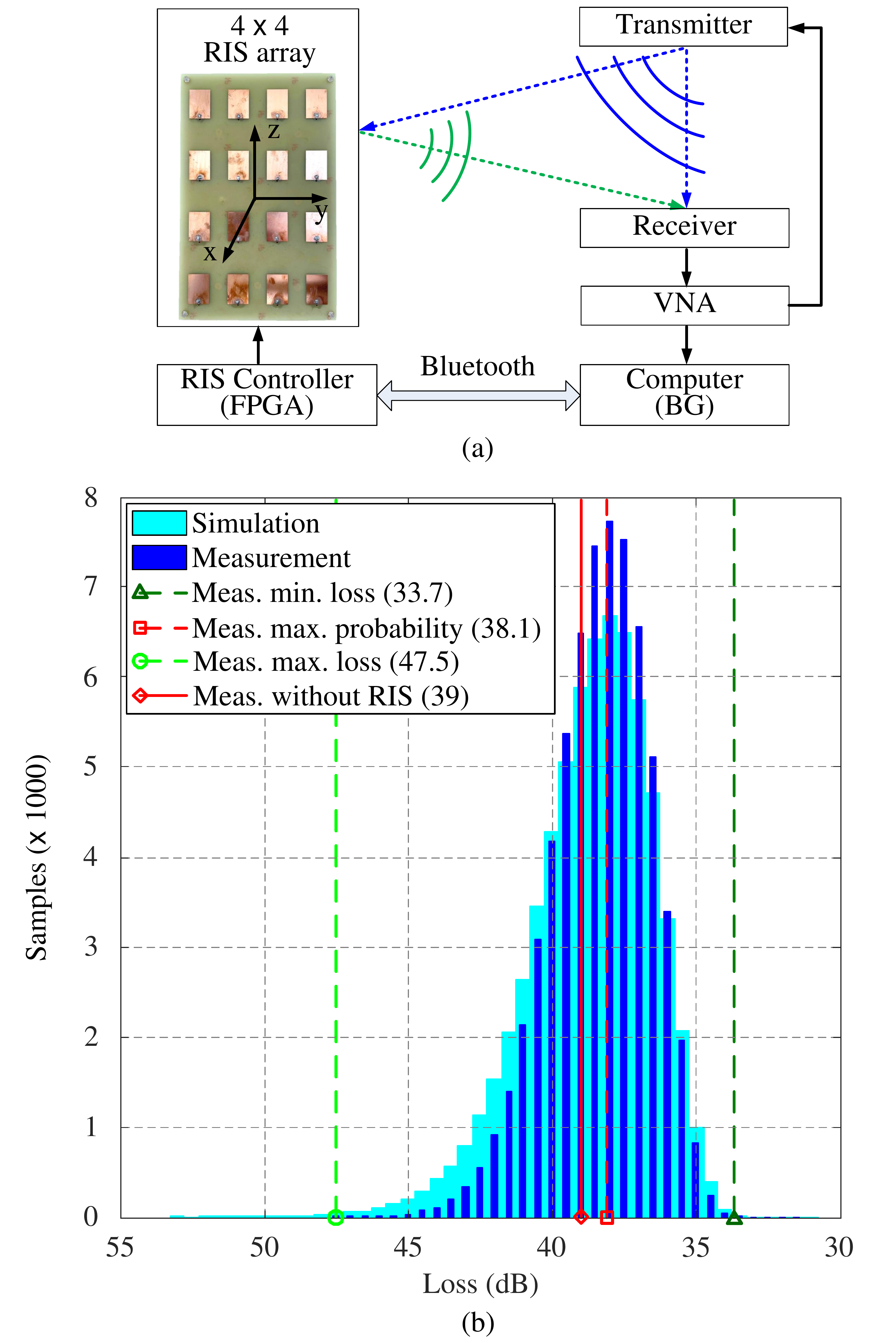} }%
    \caption{Using the $4 \times 4$ RIS array, (a) coordinate system and block diagram of experimental setup, and (b) probability mass function of simulated and measured data for exhaustive DPSs' states.
    \label{fig:4x4RIS}}
\end{figure}

Our experiments utilize a ${4 \times 4}$ RIS array, as shown in Fig. \ref{fig:4x4RIS}(a). The array is composed of $16$ RIS elements, as illustrated in Fig. \ref{fig:RISelement}. The transmitter and receiver both use a patch antenna that is identical to the antenna used on the RIS element. The received signal quality for the BG algorithm is determined by measuring the gain loss using a VNA at 3.5 GHz. We implement the BG algorithm using MATLAB on a personal computer, and the RIS controller is a DE10-Nano Kit based on a Cyclone V SoC FPGA. The controller configures all the DPSs using general-purpose input/output (GPIO) pins. The communication between the computer and the RIS controller is through a half-duplex Bluetooth connection. The computer determines the DPSs' states using the BG algorithm, and the RIS controller processes the command from the computer and sets the DPSs' states on the RIS array.

To evaluate \eqref{eq:AllArray} in real-world scenarios, we use the coordinate system and location of the $4 \times 4$ RIS array as shown in Fig. \ref{fig:4x4RIS}(a). The RIS array is located at $(0, 0, 0)$ meters, the transmitter is located at $(0.8, 0, 0)$ meters, and the receiver is located at $(0.8, 0.2, 0)$ meters. The transmitter and receiver face along the negative x-axis with the same polarization direction as the RIS array.

By testing all possible states of the DPSs, we can evaluate \eqref{eq:AllArray} by comparing the probability mass function (pmf) of the simulated and measured data. However, the number of total states is $2^{N_{\rm bit} \times N_{\rm ris}}$, which would result in an extremely long measurement time. Therefore, we only choose two states for each DPS, resulting in $2^{N_{\rm ris}} = 65,536$ samples. Specifically, the $5$-th and $8$-th states of the DPS are selected because these states have similar attenuation of reflection coefficient and about $180^{\circ}$ phase difference. The corresponding pmfs for the simulated and measured data are shown in Fig. \ref{fig:4x4RIS}(b). They are very similar in shape, although there are some differences between them. The main reasons for the differences between simulated and measured data are: First, the simulation of the SM of RIS is based on the assumption that the RIS array is a perfect planar structure of a good conductor, while in reality, the RIS array is comprised of two layers, which makes the assumption not entirely correct. Second, the radiation pattern of the RIS array is directly composed of multiple RIS elements, and the mutual coupling effect is ignored. Despite these limitations, the model of \eqref{eq:AllArray} is verified.

Another key finding from our experiments is that the gain difference between the case with the best RIS state and the case without RIS is $5.3$ dB, and the gain difference between the cases with the best and worst RIS states is $13.8$ dB. These differences demonstrate that our RIS array is able to significantly improve the received signal quality by controlling the states of the DPSs. Under the same experimental setup, when using the BG algorithm, we can achieve a loss of $34.9$ dB, which corresponds to the top $0.67\%$ of all possible states. Therefore, the BG algorithm is an efficient algorithm for finding the optimal state of DPSs. We will use the BG algorithm in the following simulations to analyze the properties of \eqref{eq:AllArray}.

\subsection*{B. Impact of Polarization on RIS Control Methods}

\begin{figure}
    \centering
    \resizebox{3.6in}{!}{%
    \includegraphics*{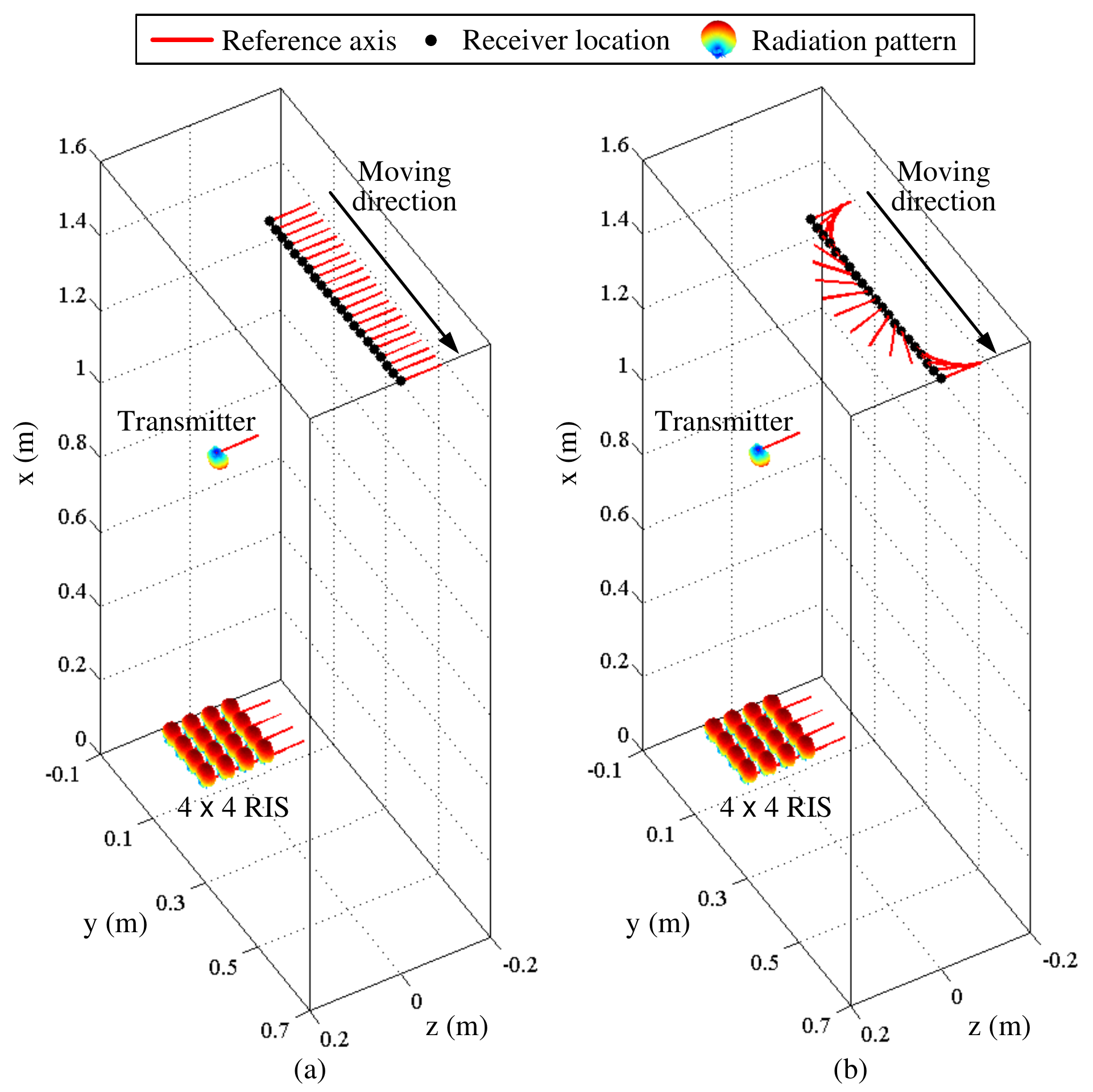} }%
    \caption{In the scenario with $4 \times 4$ RIS array, the moved receiver (a) maintains or (b) rotates its orientation.
    \label{fig:SimScenario}}
\end{figure}

The simulation setup, as shown in Fig. \ref{fig:SimScenario}, is similar to the experimental setup in Fig. \ref{fig:4x4RIS}(a), where the ${4 \times 4}$ RIS array is employed. First, we examine the impact of the polarization on the received signal at the receiver. The location of the receiver is moved from $(1.6, 0.2, 0)$ to $(1.6, 0.7, 0)$ in $0.025$\,m steps along the y-axis. The receiver's position is either maintained or rotated, as shown in Figs. \ref{fig:SimScenario}(a) and (b), respectively, with the reference axis (red line) representing the reference polarization direction. All antennas are the same patch antenna structure, and the radiation pattern of patch antenna is shown in Fig. \ref{fig:EMangle}.

Notably, even though there is a LoS path between the transmitter and receiver, the path loss is very large due to the weak gain of the radiation pattern at those AoAs and AoDs. Therefore, the path loss through the LoS is much larger than the path loss through the RIS, making the contribution from the LoS negligible. Thus, the following results are mainly contributed by the RIS.

\begin{figure}
    \centering
    \resizebox{3.6in}{!}{%
    \includegraphics*{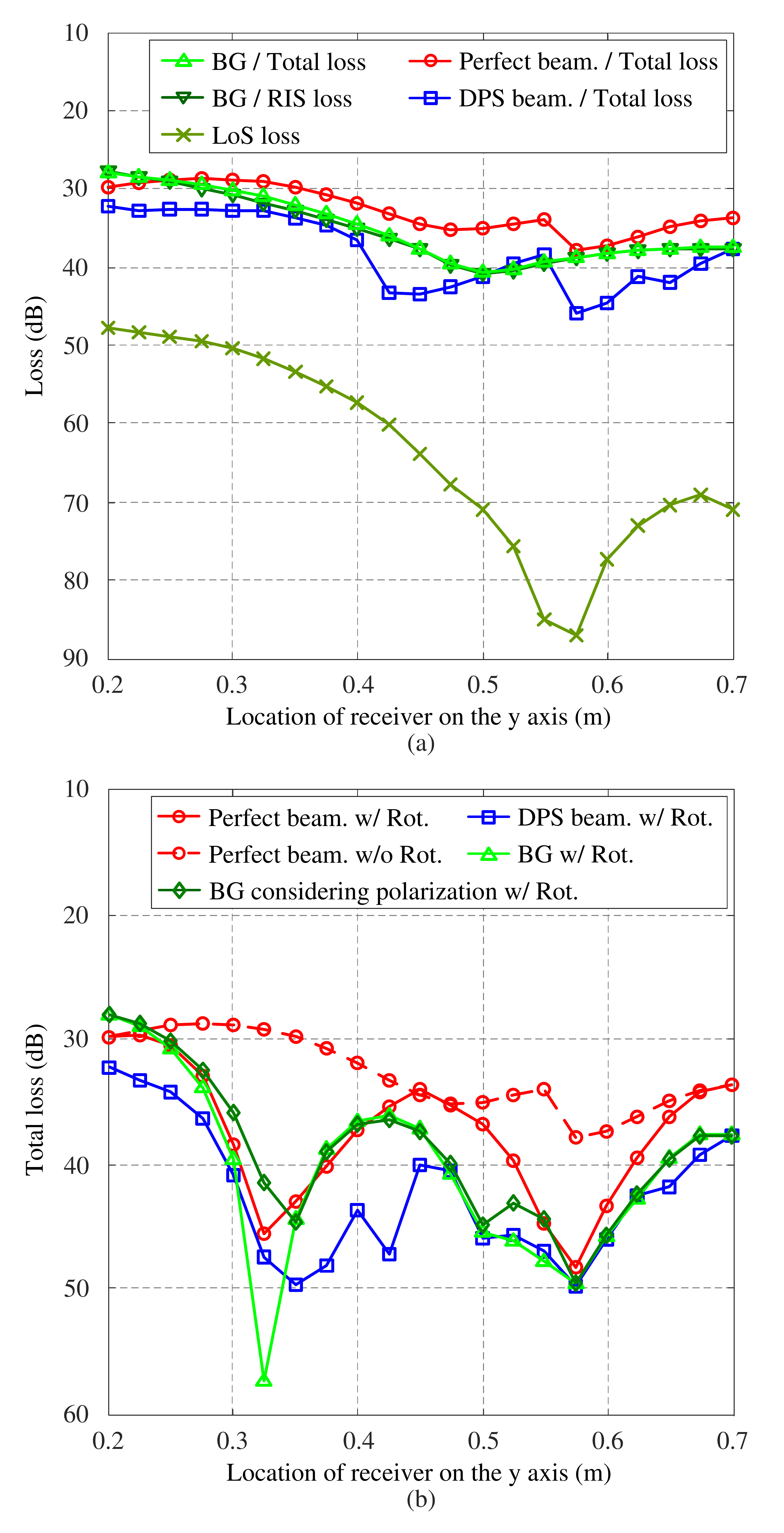} }%
    \caption{Different controlling methods for the (a) matched and (b) mismatched polarization.
    \label{fig:SimCompar}}
\end{figure}

The results corresponding to Fig. \ref{fig:SimScenario}(a) are shown in Fig. \ref{fig:SimCompar}(a). In Fig. \ref{fig:SimScenario}(a), the polarization among the antennas (i.e., transmitter, RIS, and receiver) is matched as they have the same orientation of the reference axis. Three RIS controlling methods are analyzed: the BG algorithm, perfect beamforming, and beamforming with DPS. The perfect beamforming method controls the phase of each RIS element according to the phase difference between the LoS path and the path of each RIS element. The perfect beamforming method serves as the upper bound of performance that cannot be achieved in practice as it does not take into account the phase quantization and the phase-dependent attenuation effect. The beamforming with DPS method, however, is based on perfect beamforming and selects the nearest phase according to the used DPS, but it takes into account the phase-dependent attenuation effect.
As expected, the perfect beamforming method shows the best result in Fig. \ref{fig:SimCompar}(a). However, the beamforming with DPS shows a significant degradation and performs worse than the BG algorithm due to ignoring the phase-dependent attenuation effect. This result indicates that only considering the phase alignment is insufficient to achieve optimal performance.

In Fig. \ref{fig:SimScenario}(b), the receiver's orientation changes at every step during movement. The receiver's antenna is placed on the yz-plane, and its reference axis is rotated counterclockwise around the x-axis. The polarizations among the antennas are mismatched except for when the reference axis rotates $180$ and $360$\,degrees. The corresponding results are shown in Fig. \ref{fig:SimCompar}(b). For perfect beamforming, three peaks can be observed at the beginning, middle, and end as the polarizations are matched at these points. Similarly to the previous case, when considering a practical DPS, beamforming with DPS is not always better than the BG algorithm. The results again indicate that only considering phase alignment is insufficient, whether the polarizations among antennas are matched or not.

From the above simulation, we can infer that if the polarization of the RIS is controllable, the performances are expected to be improved. A structure for two ports was proposed in \cite{Nawaz-2016}, which provides dual polarization with the same metal plate of the antenna. By using this structure, it is possible to switch the location of the port on the same metal plate of the antenna and obtain polarization switching on a RIS element. The BG algorithm can be extended to this case by including different polarizations and DPS states. The considered polarizations are linear with $0$, $45$, and $90$ degree rotations. The corresponding results are also shown in Fig. \ref{fig:SimCompar}(b). At a location of nearly $0.325$\,m, where the reference axis of the receiver is orthogonal to the RISs, the BG algorithm without polarization selection obtains the worst result. However, the BG algorithm considering polarization selection improves by about $15$ dB by choosing the better AM of RIS. This result indicates that controlling the orientation of polarization is necessary.

\begin{figure}
    \centering
    \resizebox{3.6in}{!}{%
    \includegraphics*{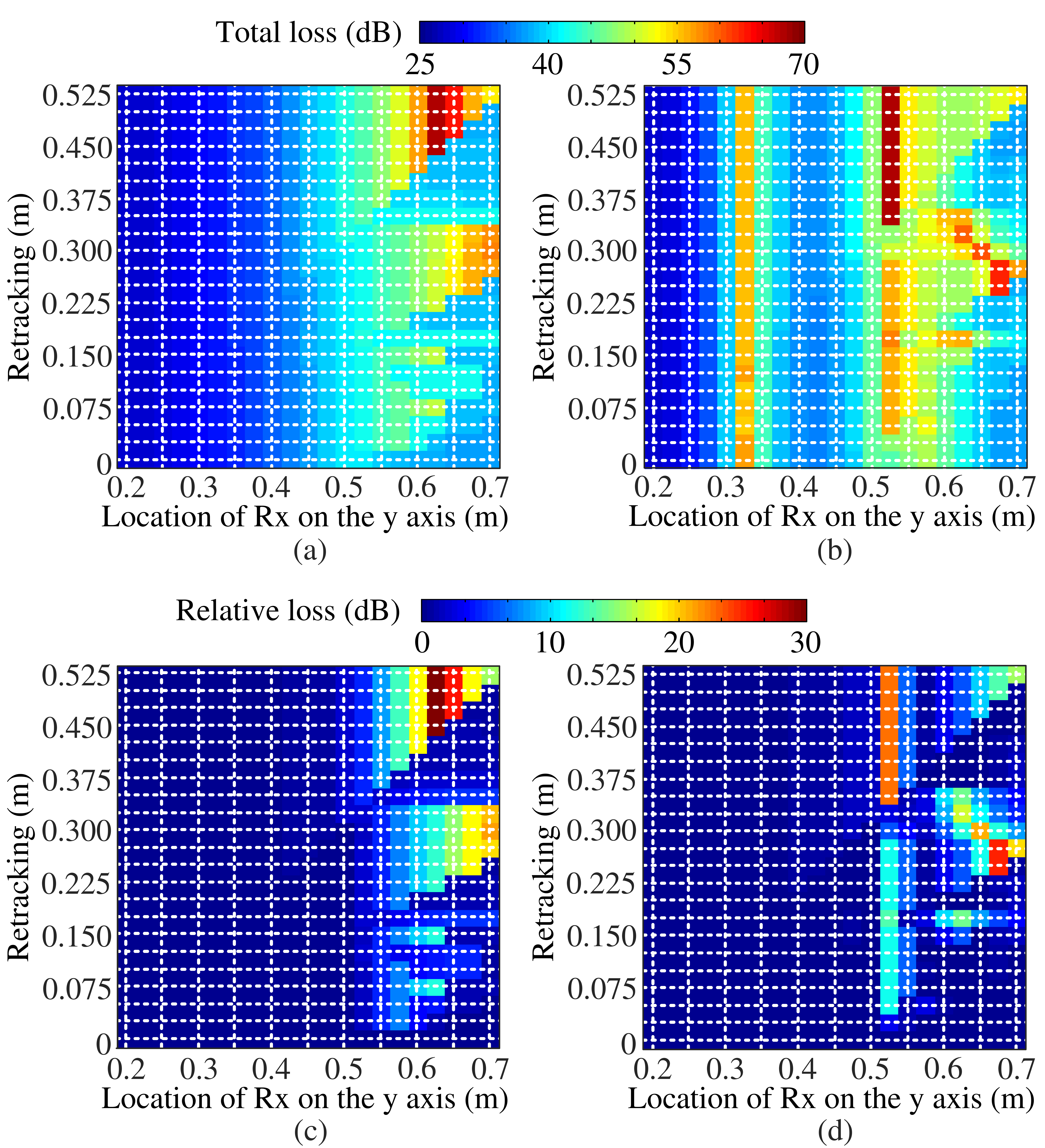} }%
    \caption{With the tracking mechanism of the BG algorithm, the total loss of (a) matched and (b) mismatched polarization, and the relative loss of (c) matched and (d) mismatched polarization.
    \label{fig:SimTrack}}
\end{figure}
\begin{figure}
    \centering
    \resizebox{3.6in}{!}{%
    \includegraphics*{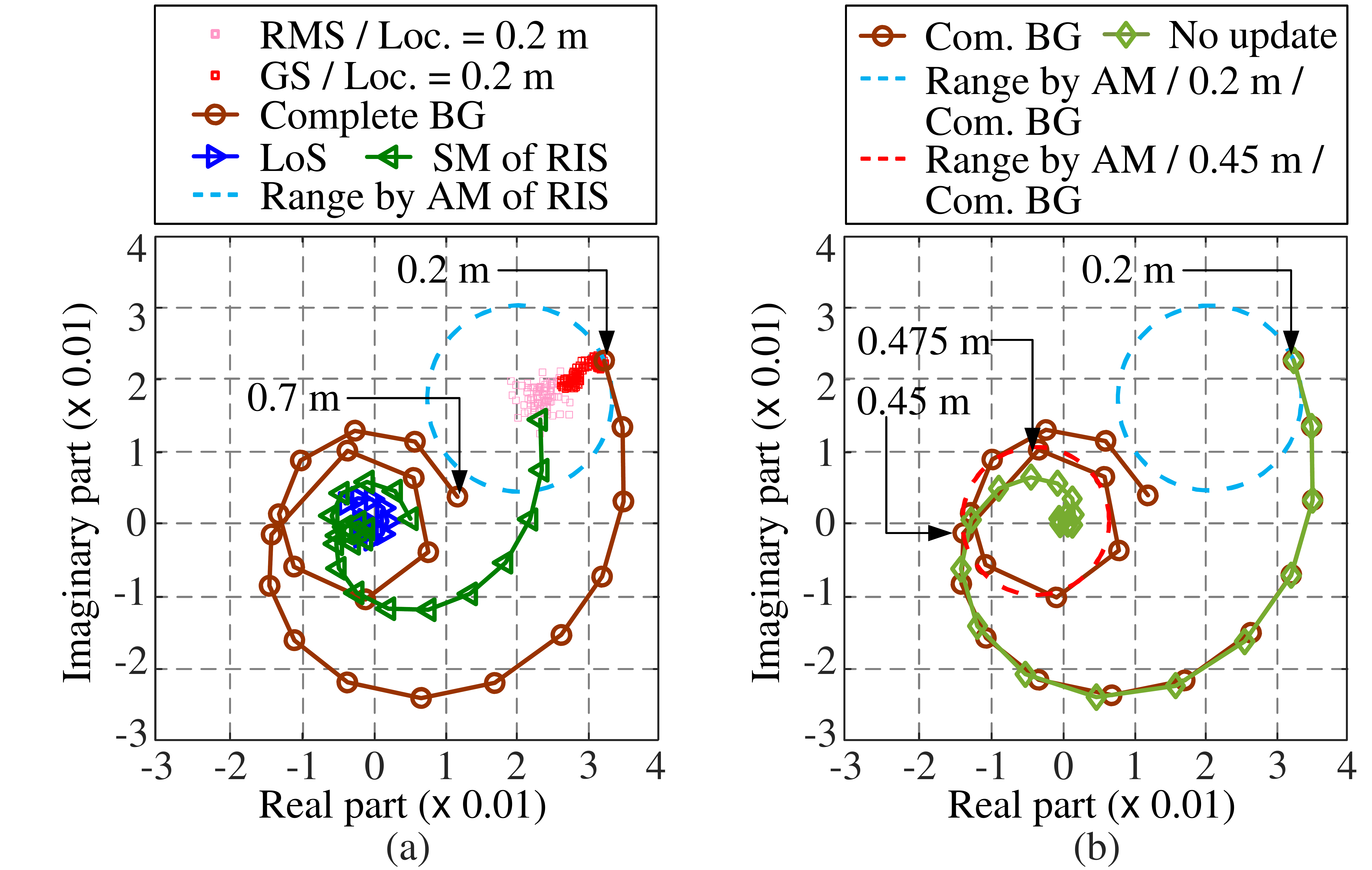} }%
    \caption{In the case of matched polarization, the channel coefficient with (a) the complete BG algorithm and (b) no update compared with (a).
    \label{fig:AnalyzeTrack}}
\end{figure}

\subsection*{C. RIS Control through Tracking Mechanism}

The existing blind controlling algorithm, as described in \cite{Ren-2022}, requires a full search to be performed each time the location changes, resulting in significant delays in practical applications. Given that the distribution of EM waves is continuous in space, it is likely that the optimal combinations of all DPSs' states in nearby locations will be similar. Based on this observation, we propose a new approach, known as the \emph{tracking mechanism}. This approach involves activating the GS component of the BG algorithm using the latest optimal combination $\qB$ as the starting state $\qB_{\rm g}$ for the new location. By only activating the GS component rather than the complete BG algorithm, the time required to search for the optimal state at the new location can be reduced.

The tracking mechanism is an effective approach to reduce the time required to search for the optimal state. However, determining when to re-activate the complete BG algorithm is a key question. To answer this, we analyzed scenarios depicted in Fig. \ref{fig:SimScenario} and plotted the total loss for the cases shown in Figs. \ref{fig:SimScenario}(a) and (b) in Fig. \ref{fig:SimTrack}(a) and (b), respectively. The horizontal axis represents the location of the receiver, and the vertical axis shows the activation distance of the tracking mechanism. Note that the receiver moves from locations $0.2$\,m to $0.7$\,m along the y-axis with a moving distance of $0.5$\,m. On the one hand, an activation distance of zero indicates that the complete BG algorithm is reactivated at every new location, resulting in the best performance when compared to any non-zero activation distances. On the other hand, an activation distance of $0.525$\,m represents no update after the BG algorithm is performed at the initial location, resulting in the worst performance.

To facilitate comparison, we also plot the relative performance loss (path loss relative to zero activation distance) of Figs. \ref{fig:SimTrack}(a) and (b) in Figs. \ref{fig:SimTrack}(c) and (d), respectively. It is found that even without performing tracking (i.e., when the activation distance is $0.525$\,m), the relative performance loss is less than $10$\,dB for receiver locations between $0.2$ and $0.5$\,m. Therefore, the tracking mechanism is less critical in this range. However, for receiver locations greater than $0.5$\,m, the activation distance for the tracking mechanism must be less than $0.225$\,m to avoid a relative loss of more than $10$\,dB.

To further understand the reason behind the above result, we analyze the channel coefficient in the case of matched polarization. Note that in \eqref{eq:AllArray}, the channel coefficient is composed of three components: the LoS path, the SM of RIS, and the AM of RIS. We depict the channel coefficient with the complete BG algorithm for each location of the receiver in Fig. \ref{fig:AnalyzeTrack}(a) by a brown line with circle markers.
Additionally, we depict the partial channel coefficient consisting only of the static components, i.e., the LoS path and SM of RIS, by a blue line with right-pointing triangle markers and a green line with left-pointing triangle markers, respectively. The sum of the static components for each location is simply called the static center point.
At the first location, i.e., $0.2$\,m, we depict the footprints of the channel coefficients during the complete BG algorithm, which includes both RMS and GS. From the figure, we observe that during the GS, the channel coefficient is the constructive interference between the AM of RIS and the sum of the static components, and thus moves gradually up and to the right. Destructive interference may occur during RMS. We can determine the controllable range from the point at static center point
to the final channel coefficient (circle marker). Because the constructive and destructive interference occur around the static center point, we can roughly determine that the shape of the controllable range is a circle. The radius of the controllable range is determined by the AM of RIS.

To compare with Fig. \ref{fig:AnalyzeTrack}(a), we now consider the worst performance when the tracking mechanism is not used, and its corresponding channel coefficients are shown in Fig. \ref{fig:AnalyzeTrack}(b) by a green line with diamond markers. Two notable phenomena are observed at receiver locations of $0.2$ and $0.45$,m. First, the radius corresponding to the AM of RIS decreases because the receiver moves away from the center of the main lobe of the RIS. Due to the decreasing radius of the controllable range, the difference between different combinations of all DPSs' states also decreases, and thus the tracking mechanism may easily select unsuitable combinations. Second, the channel coefficient changes significantly with the complete BG algorithm as the receiver moves from $0.45$ to $0.475$\,m. At locations greater than $0.45$\,m, the channel coefficient without any update differs noticeably from that obtained with the complete BG algorithm. This necessitates a lower activation distance for the tracking mechanism to mitigate the need for high tracking ability. Consequently, when the AM of RIS dominates in the channel coefficients, its tracking ability based on the continuity property of the EM wave in space is much better than in scenarios where the SM of RIS dominates. In this case, the states of RIS do not need to be updated as frequently.

\section*{VI. Conclusion}

In this paper, we have presented a new channel model for RIS that takes into account the polarization and is composed of the AM and the SM of RIS. The AM of RIS is based on the antenna resonant mode corresponding to the resonant length on the metal plate of RIS, while the SM of RIS consists of the scattered and reflected waves on the metal surface of RIS. Previous works have not clearly addressed these two effects, which limits their practical use.
We implemented a $4 \times 4$ RIS array to verify the proposed channel model and to control the AM of RIS with DPSs, where we clarified the reason of phase-dependent attenuation by presenting a model for the DPS. We also proposed the BG algorithm to control the proposed channel model for RIS, which reduces the complexity of channel acquisition. Our simulation results indicate that the polarization of the antenna, phase-dependent attenuation, and tracking mechanism are important considerations for the control algorithm. Specifically, the proposed tracking mechanism can significantly reduce the time required to search for the optimal state at the new location only when the strength of the AM of RIS dominates the channel coefficient. These findings provide insights into the design and optimization of RIS-based communication systems.


{\renewcommand{\baselinestretch}{1.1}
\begin{footnotesize}
\bibliographystyle{IEEEtran}

\end{footnotesize}}

\end{document}